\documentclass[sigconf, nonacm, prologue, table]{acmart}
\AtBeginDocument{%
  }

\setcopyright{none}
\usepackage{microtype}
\usepackage{graphicx}
\usepackage{booktabs}
\usepackage{subcaption}
\usepackage{xcolor}
\usepackage{algorithm}
\usepackage{algorithmic}
\usepackage{multirow}
\usepackage{lipsum}
\usepackage{xparse}
\usepackage{tikz}
\usepackage{hyperref}
\usepackage{dutchcal}
\usepackage{bm}
\usepackage{todonotes}
\usepackage{enumitem}
\usepackage{array}
\usepackage{soul}

\newcommand{\brackets}[1]{\textnormal{[#1]}}

\newcommand{\method}{$\nabla$-\,CMH}

\newcommand{\candlist}[1]{\tilde{A}_{#1}} 

\newcommand{\graph}{G}
\newcommand{\nodes}{V}
\newcommand{\edges}{E}

\newcommand{\f}{f(\cdot)}

\begin{document}

\title[The Right to Hide: Masking Community Affiliation via Minimal Graph Rewiring]{The Right to Hide:\\Masking Community Affiliation via Minimal Graph Rewiring}



\author{Matteo Silvestri}

\email{silvestri.m@di.uniroma1.it}
\affiliation{%
  \institution{Sapienza University of Rome}
  \city{Rome}
  \state{}
  \country{Italy}
}

\author{Edoardo Gabrielli}
\email{edoardo.gabrielli@uniroma1.it}
\affiliation{%
  \institution{Sapienza University of Rome}
  \city{Rome}
  \state{}
  \country{Italy}
}

\author{Fabrizio Silvestri}
\affiliation{%
  \institution{Sapienza University of Rome}
  \city{Rome}
  \state{}
  \country{Italy}
}

\author{Gabriele Tolomei}
\affiliation{%
  \institution{Sapienza University of Rome}
  \city{Rome}
  \state{}
  \country{Italy}
}

\renewcommand{\shortauthors}{Silvestri et al.}

\begin{abstract}
  Protecting privacy in social graphs may require obscuring nodes' membership in sensitive communities. However, doing so without significantly disrupting the underlying graph topology remains a key challenge.
  In this work, we address the \emph{community membership hiding} problem, which involves strategically modifying the graph structure to conceal a target node's affiliation with a community, regardless of the detection algorithm used.
  We reformulate the original discrete, counterfactual graph search objective as a differentiable constrained optimisation task.
  To this end, we introduce \method{}, a new gradient-based method that operates within a feasible modification budget to minimise structural changes while effectively hiding a node's community membership. 
  Extensive experiments on multiple datasets and community detection methods demonstrate that our technique outperforms existing baselines, achieving the best balance between node hiding effectiveness and graph rewiring cost, while preserving computational efficiency.
\end{abstract}

%



\maketitle

\section{Introduction}
\label{sec:introduction}
Community detection is a fundamental tool for analysing complex graph structures such as social networks, biological systems, and communication networks~\cite{fortunato2010pr}. It is typically performed using \textit{community detection algorithms}~\cite{blondel2008jsm}, which aim to uncover groups of tightly connected nodes -- called \textit{communities} -- that exhibit similar characteristics, interactions, or structural patterns.

The ability to identify such communities has enabled a wide range of applications across diverse domains~\cite{karastas2018ibigdelft, cd_applications}, including targeted advertising~\cite{mosadegh2011ajbmr}, recommendation systems~\cite{cd_recsys}, and network security~\cite{cd_networks_Sec}. However, despite its utility, community detection also raises significant privacy concerns, especially in the context of social graphs, where it can inadvertently reveal sensitive affiliations or personal traits. For example, community assignments may expose users' political or religious beliefs, age, gender~\cite{waller2021nature}, or ties to controversial or conspiratorial groups~\cite{samory2018aaai}.

While one option for protecting privacy is to leave the platform entirely, such an action is often too drastic. A more flexible approach would enable individuals to control their visibility within detected communities while continuing to participate. This strikes a better balance between preserving privacy and maintaining the utility of community detection. Furthermore, it better aligns with modern data protection initiatives such as the European GDPR~\cite{regulation2016general} and AI Act \cite{act2024eu}, which includes provisions like the ``\textit{right to be forgotten}''~\cite{TRTBF2013}. 

Motivated by this challenge, we address the problem of \textit{community membership hiding}, first introduced by~\citet{bernini2024kdd}. This task draws inspiration from \textit{counterfactual reasoning} \cite{tolomei2021tkde, tolomei2017kdd,lucic2022aistats}, and involves strategically modifying a graph's topology to prevent a target node from being identified as part of a specific community by detection algorithms.

In this work, we build upon the method proposed by~\citet{bernini2024kdd}, formulating the community membership hiding task as a counterfactual graph objective. Specifically, we cast it as a constrained optimisation problem, where the goal is to perturb the structure surrounding the target node to obscure its community affiliation. However, unlike the original approach that treats the problem as a discrete objective and solves it via deep reinforcement learning, our method is inspired by adversarial attacks on graph neural networks~\cite{trappolini2023savage}. We reformulate the task as a \textit{differentiable} objective, enabling efficient solution through gradient-based optimisation techniques. This continuous formulation allows for minimal structural modifications while achieving effective concealment.
Our approach offers three key advantages over the original technique~\cite{bernini2024kdd}: $(i)$ a higher success rate in the community membership hiding task, $(ii)$ meticulous allocation of the available budget to achieve the goal, and $(iii)$ improved computational efficiency.

Our main contributions are summarised below.

\begin{enumerate}[label=(\arabic*), itemsep = 4pt, topsep=5pt]
    \item We reformulate the original community membership hiding task as a differentiable counterfactual graph objective;
    \item We propose \method{}, a gradient-based method that strategically perturbs the target node's neighbourhood to hide;
    \item We evaluate our approach on real-world graph datasets and show its superiority over existing baselines. The source code of our method is available at: {\url{https://anonymous.4open.science/r/community-membership-hiding-B188/}}.
\end{enumerate}

The remainder of the paper is organised as follows. Section~\ref{sec:related} reviews related work. In Section~\ref{sec:background}, we present background and preliminaries. Section~\ref{sec:problem} reformulates the problem setting. Our proposed method is detailed in Section~\ref{sec:method}, followed by an extensive empirical evaluation in Section~\ref{sec:experiments}. We discuss the current limitations of our method in Section~\ref{sec:limitations}. Finally, Section~\ref{sec:conclusion} concludes the paper.

\section{Related Work}
\label{sec:related}
The body of related work primarily falls into two key areas: \textit{community detection} and \textit{community membership hiding}. The latter also shares connections with \textit{adversarial attacks on graphs}. Below, we review the most relevant contributions in each of these domains. 

\smallskip
\noindent \textbf{\textit{Community Detection.}} Community detection algorithms play a crucial role in analysing graph structures by identifying and grouping nodes into \textit{communities}. These communities are clusters of nodes that exhibit a higher density of connections within the group compared to their connections with the rest of the graph. 
Existing approaches to identify non-overlapping communities include Modularity Optimisation~\cite{greedy_detection_alg, louvain_detection_alg, leiden}, Spectrum Optimisation~\cite{spectrum_analysis}, Random Walk~\cite{walktrap_detection_alg}, Label Propagation~\cite{label_detection_alg}, or Statistical Inference~\cite{mmsbm_detection_alg}. 
In contrast, overlapping community detection algorithms frequently use methods such as Matrix Factorisation~\cite{bigclam_detection_alg}, Neighbourhood-Inflated Seed Expansion~\cite{nise}, or techniques based on minimising the Hamiltonian of the Potts model~\cite{Ronhovde_2009}. For a comprehensive overview of these methods, see the extensive summary by~\citet{community_detection_survey}. Furthermore, deep learning models have been increasingly employed to tackle the complex problem of community detection~\cite{deepl_detection_alg, deepl_procd}, with DGCLUSTER~\cite{deepl_dgcluster} representing a notable approach.

\smallskip
\noindent \textbf{\textit{Community Membership Hiding.}} Community membership hiding addresses the problem of concealing a single node's affiliation with a particular community. 
The seminal work in this area is by~\citet{bernini2024kdd}, who introduce \textit{DRL-Agent}, a method that strategically modifies a node's local neighbourhood to obscure its community membership. They formulate the task as a counterfactual graph objective and leverage a graph neural network (GNN) to capture the input graph's structural complexity. Their approach uses deep reinforcement learning (DRL) within a Markov decision process to determine which edges to modify, under a fixed budget to limit the number of changes for efficiency and realism. 
\\
Unlike DRL-Agent, our method reformulates the problem as a \textit{differentiable} counterfactual objective, enabling the use of well-known gradient-based optimisation techniques. This brings three key benefits: $(i)$ improved node hiding effectiveness, $(ii)$ a more efficient use of the available budget, avoiding its full exhaustion, and $(iii)$ lower computational overhead by avoiding costly DRL training.

\smallskip
\noindent \textbf{\textit{Adversarial Attacks on Graphs.}} 
The task of hiding a node's community affiliation can also be interpreted as a targeted objective within the broader framework of adversarial attacks on graphs.
Although this domain has been extensively explored~\cite{attack_graph_survey1, attack_graph_survey2}, most efforts focus on evading link prediction~\cite{trappolini2023savage, attack_graph_link1, attack_graph_link2}, or disrupting node and graph classification~\cite{attack_graph_node, attack_graph_graph}, leaving attacks against non-parametric graph clustering unexplored~\cite{cluster_attack}. In addition, many methods target specifically GNNs, limiting their applicability.
\\
Some works tailored for community detection aim to hide groups of nodes by dispersing them across communities~\cite{deception_modularity_2, deception_modularity_3, cdattack}. These approaches focus on group-level obfuscation and rely on global modifications, which are ill-suited for scenarios requiring localised changes.
Other works, in contrast, aim to disrupt community detection performance by minimising modularity~\cite{q_attack} or altering node-level features, such as centrality~\cite{deception_modularity_2}.
However, we tackle a more fine-grained problem: obscuring the community membership of a single target node by modifying \emph{only} its local neighbourhood -- namely, by altering edges that the target node itself can control.

\section{Background and Preliminaries}
\label{sec:background}

In this section, we first introduce the notation used throughout the paper. We then provide a brief overview of the well-known community detection problem, which forms the foundation for defining the community membership hiding problem.


Let $\graph =(\nodes,\edges)$ be an arbitrary (undirected\footnote{The same reasoning easily extends to the case where $\graph$ is directed.}) graph, where $\nodes$ is the set of nodes with $|\nodes|=n$, and $\edges \subseteq \nodes \times \nodes$ is the set of edges with $|\edges|=m$. The structure of $\graph$ is represented by a binary adjacency matrix denoted by $A = (A_{u,v})_{u,v \in \nodes}$, where $A_{u,v} = 1$ if there is an edge between nodes $u$ and $v$, i.e., $(u,v) \in \edges$, and $A_{u,v} = 0$ otherwise. 
The neighbourhood of a node $u$, defined as the set of nodes directly connected to $u$, corresponds to the $u$-th row of $A$. We denote this row as $A_u$ and refer to it as the \emph{adjacency vector} of $u$. 


The \emph{community detection} problem aims to partition the nodes of a graph into clusters, referred to as \emph{communities}. Intuitively, communities are groups of nodes with strong intra-cluster connections compared to their links with nodes outside the cluster. In this work, we focus on detecting non-overlapping communities based solely on the graph's edge structure, as in \cite{bernini2024kdd}, leaving the exploration of methods that consider node features to future research.
Formally, a community detection algorithm is a function $f(\cdot)$ that partitions the graph $\graph$ into a set of non-empty, disjoint communities $f(\graph) = \{C_1,C_2,\dotsc,C_k\}$, where each node $u$ is assigned to \emph{exactly one} community, and the number of communities $k$ is typically unknown.
Note that the actual input to $f$ can be \textit{any} suitable representation of $\graph$, ranging from the simple adjacency matrix $A$ to more complex forms such as $\mathcal{g}=(A,X)$, where $X$ is the node feature matrix and $\mathcal{g}$ is a graph neural network. Without loss of generality, hereinafter we denote the input simply as $\graph$.

Community detection algorithms often aim to maximise a score that quantifies intra-community cohesiveness. A widely used metric for this purpose is Modularity~\cite{newman2006pnas}. However, optimising Modularity is generally NP-hard. To address this challenge, numerous practical approximation methods have been developed. Notable examples include Greedy~\cite{greedy_detection_alg}, Louvain~\cite{louvain_detection_alg}, Leiden~\cite{leiden}, WalkTrap~\cite{walktrap_detection_alg}, InfoMap~\cite{infomap_detection_alg}, Label Propagation~\cite{label_detection_alg}, Leading Eigenvectors~\cite{eigenvectors_detection_alg}, Edge-Betweenness~\cite{edge_detection_alg}, and SpinGlass~\cite{spinglass_detection_alg}.

\section{Community Membership Hiding (CMH)}
\label{sec:problem}

\begin{figure*}[t]
    \centering
    \subfloat[Target node $u$ and the communities $f(\graph)$ detected by the \textit{Louvain} algorithm.]{%
        \includegraphics[width=0.3\linewidth]{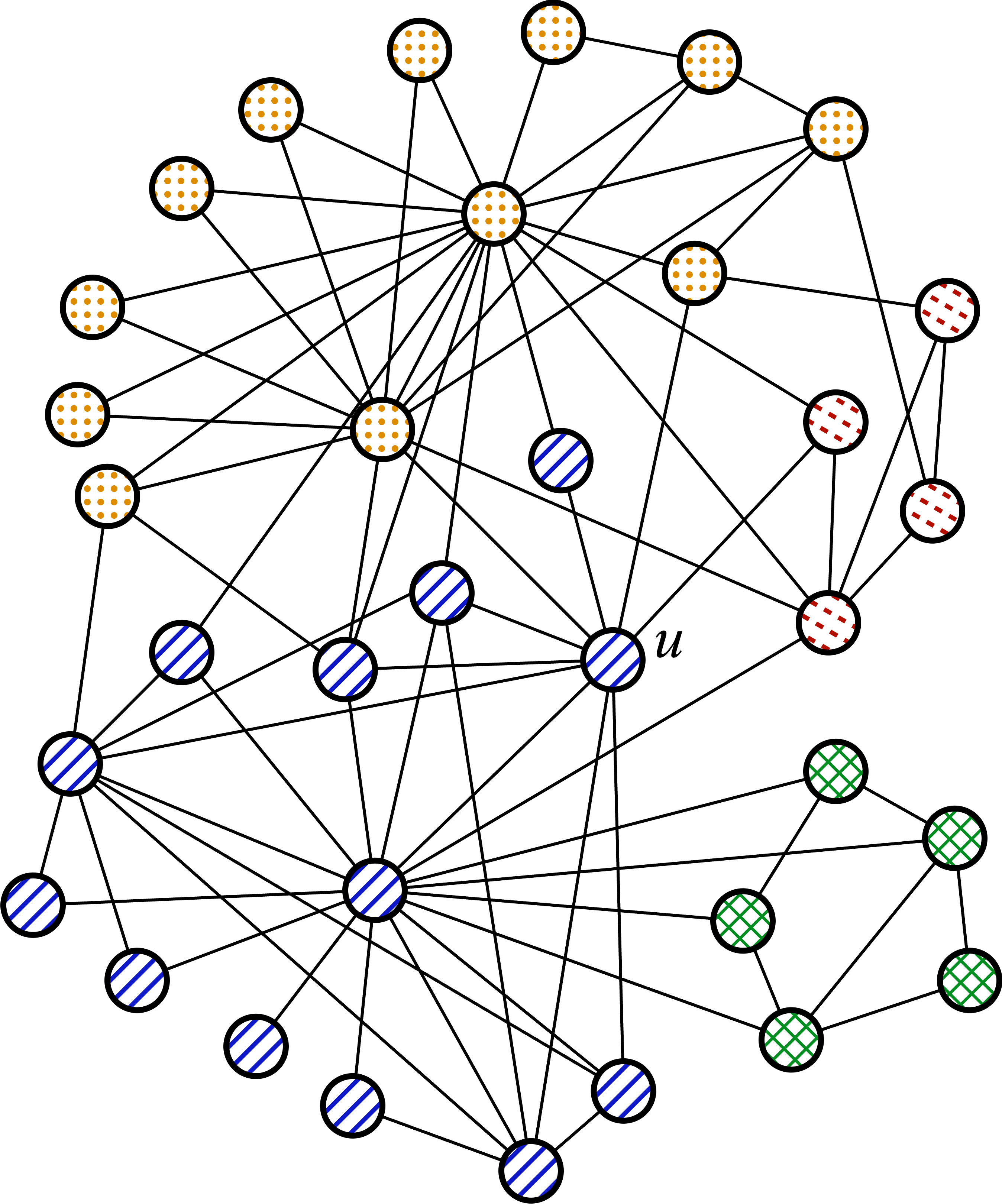}%
        \label{fig:old_comm}}
    \hfill
    \subfloat[Counterfactual graph $\graph'$, where bold (red) edges represent deletions and dashed (green) edges indicate additions.]{%
        \includegraphics[width=0.3\linewidth]{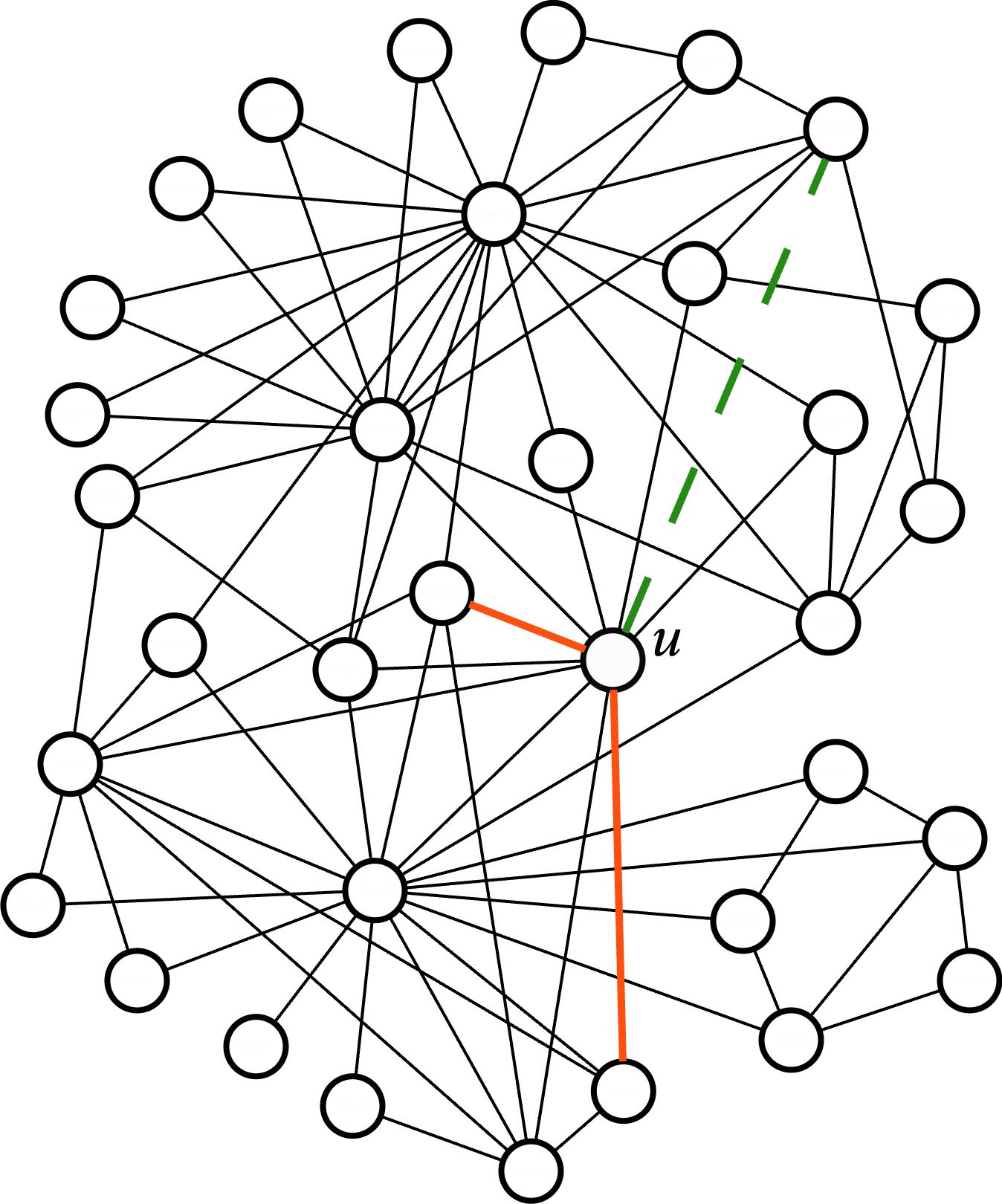}%
        \label{fig:new_graph}}
    \hfill
    \subfloat[New communities $f(\graph')$ detected on the new graph $\graph'$. The hiding objective is achieved if $sim(C_i \setminus \{u\}, C'_i \setminus \{u\})  \leq \tau$.]{%
        \includegraphics[width=0.3\linewidth]{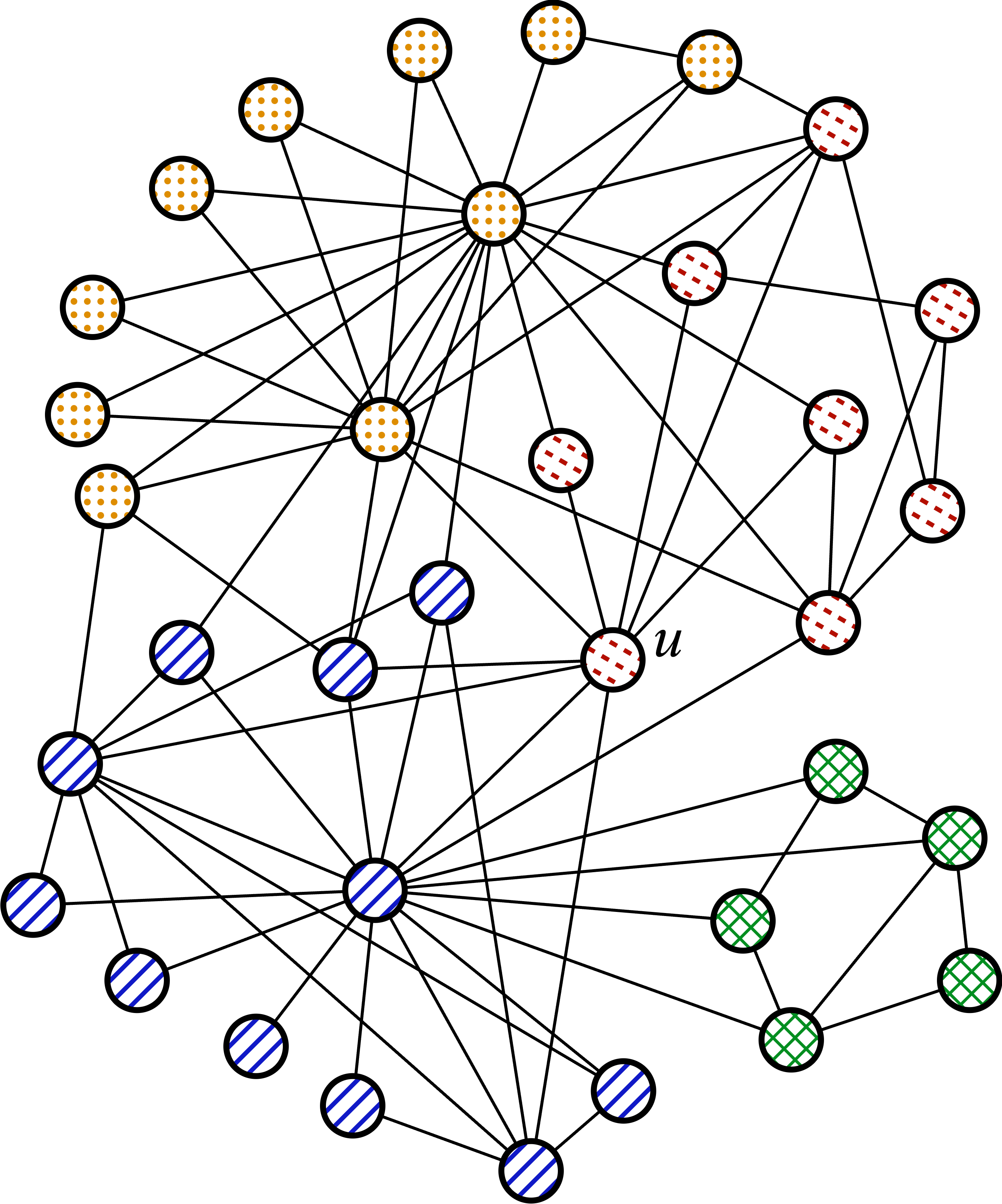}%
        \label{fig:new_comm}}
    \caption{Example of the Community Membership Hiding Problem on the Karate Club graph.}
    \label{fig:cmh_example}
    \vspace{-3mm}
\end{figure*}


In its most general form, \textit{community membership hiding} aims to prevent a specific node from being identified as part of a designated cluster by a link-based community detection algorithm. This is achieved by strategically modifying the node's connections -- i.e., its local neighbourhood -- which corresponds to altering its row in the adjacency matrix.

This definition is domain-agnostic and applies to \textit{any} graph, as reflected in the problem formulation in Section \ref{subsec:problem_Def}. 
However, the most typical scenario arises in the context of social networks. In such settings, the motivating use case involves a user who is either aware or suspects that they belong to a particular community, yet wishes to obscure this membership for privacy reasons. 

Depending on who performs the hiding, we distinguish between a \textit{platform-mediated} and a \textit{user-initiated} model. In the former, the network owner offers a hiding service, with full access to the graph and knowledge of the community detection algorithm $\f$. In the latter, the user acts independently, relying solely on access to their local neighbourhood and without knowledge of $\f$.

In this work, we focus on a \textit{restricted} platform-mediated setting, following the assumptions made by~\citet{bernini2024kdd}. Specifically, we treat $\f$ as a black box, leveraging only on its input–output behaviour without requiring insight into its internal mechanics. Note that this is a conservative assumption, since in platform-mediated scenarios $\f$ may in fact be fully known. We also assume full access to the graph topology, consistent with the case where the platform executes the hiding on behalf of the user, though our approach can be extended to partial-knowledge settings.




\subsection{Problem Definition}
\label{subsec:problem_Def}
Let $\graph = (\nodes,\edges)$ be a graph and $f(\graph) = \{C_1 \dots C_k\}$ denote the community partitioning obtained by applying a community detection algorithm $\f$. Suppose that $\f$ identifies the target node $u$ as a member of the community $C_i$, denoted as $u \in C_i$.

The goal of the community membership hiding problem, as introduced by~\citet{bernini2024kdd}, is to learn a \textit{perturbation function} $h_{\theta}(\cdot)$, parameterised by $\theta$, that transforms the original graph $\graph = (\nodes, \edges)$ into a perturbed version $\graph' = h_{\theta}(\graph) = (\nodes, \edges')$. The objective is to find a perturbation that ensures the target node $u$ is no longer assigned to its original community $C_i$ when the community detection algorithm $\f$ is applied to $\graph'$.
 
The definition of community membership hiding may vary. For instance, if a target node is reassigned to a new community $C'_i$ in the perturbed graph $\graph'$, one goal may be to minimise the similarity between the original community $C_i$ and the new community $C'_i$. Alternatively, the hiding objective might focus on ensuring that specific nodes from $C_i$ do not belong to $C'_i$. For example, if $C_i = \{ u,v,w,y,z \}$, we might aim to reassign $u$ to $C'_i$ such that $y,z \notin C'_i$. 

In this work, we adopt the first definition of membership hiding, leaving the exploration of alternative definitions for future research. Specifically, given a similarity function  $sim(\cdot,\cdot)$ and a non-negative threshold $\tau$, we consider the hiding task successful if the condition $sim(C_i \setminus \{u\}, C'_i \setminus \{u\})  \leq \tau$ is satisfied. We assume that $sim(\cdot,\cdot)$ outputs values in the range $[0,1]$ and $\tau \in [0,1)$. When $ \tau = 0 $, the condition is most restrictive, requiring no overlap between $C_i$ and $C'_i$ (excluding the node $u$). As $\tau$ increases, the condition becomes less stringent, making it easier to achieve the hiding goal.
 
More formally, the community membership hiding task resorts to solving the following constrained optimisation problem:
\begin{equation}
\label{eq:objective}
\begin{split}
    \theta^* & = \underset{\theta}{\text{arg min}} \bigg \lbrace \mathcal{L}(h_{\theta};\graph, f, u) \bigg \rbrace \\
    & \text{subject to: }  \|\bm{b}_u\|_0 \leq \beta,
\end{split}
\end{equation}
where $\mathcal{L}(\cdot)$ is a loss function defined as follows:
\begin{equation}
    \small
    \mathcal{L}(h_{\theta}; \graph,f,u) = \ell_{\textit{hide}}(h_{\theta},\graph;f,u) + \lambda \ \ell_{\textit{dist}}(h_{\theta}(\graph),\graph;f).
\end{equation}
The first term ($\ell_{\text{hide}}$) incentivises the target node $u$ to detach from its original community, i.e., to reach the specific hiding goal. The second term ($\ell_{\text{dist}}$) quantifies the impact of modifications -- e.g., by computing the distance between the original graph $\graph$ and the modified graph $\graph'$, the distance between their corresponding partitions $f(\graph)$ and $f(\graph')$, or a convex combination of both, as done by~\citet{bernini2024kdd}. The distance between partitions captures how local perturbations can alter the global community structure, potentially reshaping the memberships of nodes beyond the target, as illustrated in Fig.~\ref{fig:cmh_example}. By penalising significant alterations via the weighting factor $\lambda$, the loss function encourages minimal modifications to the graph while still attaining the desired hiding effect.

Concretely, given a fixed budget $\beta > 0$ for modifying the target node $u$'s neighbourhood, the problem reduces to identifying the optimal function $h^* = h_{\theta^*}$, where the parameters $\theta^*$ are determined by solving the constrained objective defined in Eq.~\eqref{eq:objective}. 
Here, $\bm{b}_u$ is a $|\nodes|$-dimensional binary vector such that $\bm{b}_u[v] = 1$ if and only if the edge $(u, v)$ is modified by $h_{\theta}$, and $\bm{b}_u[v] = 0$ otherwise.
Thus, Eq.~\eqref{eq:objective} can be interpreted as identifying the \textit{counterfactual graph} $\graph^* = h^*(\graph)$, i.e., a modified version of the original graph that masks the community membership of the target node $u$ when input back to the community detection algorithm $\f$.

However, in contrast to the original method~\cite{bernini2024kdd}, we adopt a more flexible strategy by not limiting the set of edges eligible for modification, i.e., any $\bm{b}_u \in \{0,1\}^{|\nodes|}$ is theoretically admissible. Specifically, we consider all possible edges between the target node $u$ and every other node in the graph, enabling both the addition and removal of connections. This unrestricted approach allows the optimisation process to be fully guided by the loss landscape, empowering it to identify the most impactful modifications.

Note that the constraint on $\|\bm{b}_u\|_0$ makes the objective non-convex, due to the inherent non-convexity of the $L^0$-norm itself. 
To overcome this challenge, rather than adopting a reinforcement learning framework as in~\citet{bernini2024kdd}, we propose a novel differentiable loss function. This formulation enables the use of efficient gradient-based optimisation methods for solving the community membership hiding task. A detailed discussion on this matter follows in the next section.

\section{Continuous Relaxation of the CMH Problem}
\label{sec:method}
The community membership hiding problem outlined in Section~\ref{sec:problem} is inherently discrete, rendering it unsuitable for direct optimisation using gradient-based techniques.
To overcome this, we adopt a strategy inspired by~\citet{trappolini2023savage} by introducing a \textit{perturbation vector} $p$ that is applied to the adjacency vector $A_u$: 
\begin{equation}
    A'_u = \operatorname{clamp}(A_u + p),
\end{equation}
where $p \in \{-1,0,1\}^{|V|}$. Intuitively, a value of $-1$ in $p$ corresponds to removing an existing edge or leaving a non-existent edge unaltered, $0$ preserves the current edge state, and $1$ either adds a new edge or retains an existing one. The function $\operatorname{clamp}(x) = \max (0, \min (x,1))$ ensures that the elements of new adjacency vector $A'_u$ are contained to $\{0,1\}$, mapping the set $\{-1,0,1,2\}$ to binary values.
 
However, since the values in $p$ remain discrete, we first introduce a real-valued vector $\hat{p}$, whose entries are constrained to the range $ [-1,1]$ using a $\tanh$ transformation. These values are then thresholded to obtain the discrete perturbation vector $p$, defined as:
\begin{equation}
    p_{i} =
        \begin{cases} 
        +1 & \text{if } \hat{p}_{i} \geq t^+, \\
        -1 & \text{if } \hat{p}_{i} \leq t^-, \\
        0 & \text{otherwise.}
        \end{cases}
\end{equation}
A straightforward choice for the thresholds is $t^+=0.5$ and $t^-=-0.5$, which cleanly separates positive, negative, and neutral perturbations. 
Consequently, $\hat{p}$ becomes the \emph{only} set of parameters subject to optimisation, fully governing the perturbation process.


\subsection{Designing a Differentiable Loss}
\label{subsec:loss}
The optimisation operates on a 
vector $\hat{p}$ initialised uniformly within $[-0.5, 0.5]^{|\nodes|}$, which corresponds to starting the process from a null perturbation state. 
Since we assume no internal knowledge of $\f$, its outcomes cannot be directly incorporated into the loss to guide optimisation. To address this, we introduce a vector $\candlist{u}$, representing what we refer to as \emph{promising actions} -- that is, edge modifications that node $u$ should prioritise to escape its current community.
In principle, as discussed in Section~\ref{subsec:problem_Def}, our framework is general enough to treat \textit{any} edge between the target node $u$ and any other node as a valid candidate for addition or removal -- i.e., any $\bm{b}_u \in \{0,1\}^{|\nodes|}$ is theoretically admissible.
However, prior work has shown that certain edge modifications are more influential than others in disrupting community assignments~\cite{deception_modularity_3, bernini2024kdd, deception_modularity_2}. Our promising actions aim to capture this intuition.

The simplest form of $\candlist{u}$ is given by $\lnot A_u$, the complement of node 
$u$'s adjacency vector. This suggests disconnecting from all current neighbours while linking to previously unconnected nodes. A more refined heuristic, which accounts for the structural properties of nodes in the graph, is presented in Section~\ref{subsec:candlist}. Moreover, the flexibility of our framework allows for the integration of more sophisticated approaches, such as \textit{learning} promising actions dynamically rather than statically defining them a priori. We leave the exploration of such adaptive strategies to future work.

Therefore, we define the first term of the loss ($\ell_{\textit{hide}}$) as:
\begin{equation}
\label{eq:l-hide}
    \small
    \ell_{\textit{hide}}(\hat{p}; A_u, \candlist{u}, q) = \| \candlist{u} - (A_u + \hat{p})\|_q,
\end{equation}
where $q \geq 1$.
In contrast, the second component of the loss ($\ell_{\textit{dist}}$) is designed to discourage large perturbations, aiming to identify the minimal counterfactual graph that causes $u$ to belong to a different community, according to $\f$.
To this end, we assess the distance between the original and intermediate adjacency vectors:
\begin{equation}
\label{eq:l-dist}
    \ell_{\textit{dist}}(\hat{p}; A_u, q) = \| A_u - (A_u + \hat{p}) \|_q = \| \hat{p} \|_q,
\end{equation}
where $q \geq 1$. 
Eventually, the objective is to determine the optimal perturbation vector $p^*$ that balances hiding effectiveness with minimal graph modifications. This formulation leads to the following constrained optimisation problem, which can now be solved via standard gradient-based methods:
\begin{equation}
\label{eq:p_objective}
\begin{split} 
    p^* & = \underset{\hat{p}}{\text{arg min}} \bigg \lbrace  \mathcal{L}(\hat{p}; A_u, \candlist{u}, q) \bigg \rbrace \\
    & \text{subject to: } \|\bm{b}_{u}\|_0 \leq \beta.
\end{split}
\end{equation}


\subsection{Promising Actions}
\label{subsec:candlist}
As discussed in Section~\ref{subsec:loss}, we employ $\candlist{u}$ as a surrogate for the output of $\f$, which cannot be directly incorporated into the loss. Since $\candlist{u}$ prioritises certain promising actions, there are various ways to define it. In this work, we adopt the following heuristic. 
Specifically, we introduce the notion of a node's significance by assigning each node $v$ a real-valued score $S_v \in [0,1]$. Accordingly, each entry of $\candlist{u}$ is defined as:
\begin{equation}
\label{eq:candlist}
        \candlist{u,v} = 
        \begin{cases}
            (1-S_v)/2 & \text{if } v \in C_i,  \\
            (1+S_v)/2 & \text{if } v \notin C_i.    
        \end{cases}
\end{equation} 
If a node $v$ belongs to the same community as $u$ ($v \in C_i$) and has a high score ($S_v \approx 1$), then $\candlist{u,v} \approx 0$, which encourages the algorithm to disconnect from it if an edge exists. Conversely, if the node lies outside $C_i$ and also has a high score, $\candlist{u,v} \approx 1$, the algorithm is more likely to establish a connection if none exists. This reflects the aim of reducing cohesiveness within the community while strengthening connections outside of it. On the other hand, when a node has a low score ($S_v \approx 0$), $\candlist{u,v} \approx \frac{1}{2}$ in both scenarios, indicating no preference for adding or removing that connection, thus favouring no changes.

For each node, we calculate the values of $K$ structural properties, denoted by $\Omega=\{\omega_1\dots,\omega_K\}$. Then, we compute a $|V|$-dimensional ranking vector $\bm{r}_{i}$ for each property $\omega_i \in \Omega$. Each element of this vector indicates the position of a node in the list of values for $\omega_i$, sorted in non-decreasing order. For example, consider a set of nodes $\nodes = \{v_1,v_2,v_3\}$ and a property $\omega_i$, with values \brackets{42,120,5}, i.e., $\omega_i(v_1) = 42$, $\omega_i(v_2) = 120$, and $\omega_i(v_3) = 5$. Sorting these values in non-decreasing order yields \brackets{5,42,120}. The ranking vector $\bm{r}_{i}$ assigns to each node the index (starting from 1) of its property value $\omega_i$ in the sorted list, resulting in $\bm{r}_{i} = $\brackets{2,3,1}, i.e., $\bm{r}_{i}[v_1] = 2$, as $\omega_i(v_1) = 42$ is the second element of the sorted list, and so on.
\\
Thus, we normalise the rankings as follows:
\begin{equation}
    S_v^{i} = \dfrac{\bm{r}_{i}[v] - 1}{|\nodes| - 1} \quad \forall v \in \nodes, \forall i=1\dots K.
\end{equation}
The final scores are obtained by aggregating the individual scores associated with each property, for example through a linear combination: $S_v = \sum_{i=1}^K  a_i \, S_v^{i} \ \forall v \in \nodes$, where $a_i \in [0,1]$ and $\sum_{i=1}^K a_i = 1$.

In this work, we consider the following structural properties of a node: $\Omega=\{$\textit{degree}, \textit{betweenness centrality}, \textit{intra/inter-community degree}$\}$. These properties are chosen to ensure consistency with the baselines methods, which also rely on them, as detailed in Section~\ref{sec:exp-setup}.
We will explore alternative structural metrics and more advanced aggregation strategies to compute $S_v$ in future work.


\subsection{\method{}}
\label{subsec:alg}
In this section, we describe our proposed method, referred to as \method{}. The method is outlined in Algorithm~\ref{alg:pseudocode},  which provides an overview of its operational mechanics.

Our approach distinguishes itself from prior work~\cite{bernini2024kdd}, which performs the maximum permitted actions without verifying the outcome of $\f$. In contrast, our technique adopts a more efficient utilisation of the available budget by dynamically recalculating the community structure after each modification to the graph. Specifically, every time $A'_u$ is altered, we reevaluate the community structure. This recalibration allows us to potentially achieve the hiding objective \textit{before} fully exhausting the allocated budget.
Furthermore, if the method depletes the budget without achieving the hiding objective, the optimisation process is restarted to explore alternative counterfactuals. This iterative restart mechanism is regulated by a predefined maximum iteration limit, denoted as $T$, which ensures computational feasibility by preventing infinite loops.

\begin{algorithm}
    \caption{\method{}}\label{alg:pseudocode} 
    \begin{algorithmic}[1]
        \REQUIRE Graph $\graph = (\nodes, \edges)$;
        target node $u$;
        community detection algorithm $f(\cdot)$; max iterations $T$;
        learning rate $\eta$;
        similarity function $sim(\cdot)$;
        budget $\beta$;
        similarity threshold $\tau$.
        \ENSURE Counterfactual graph $\graph'$
        \STATE $\hat{p} \sim \mathcal{U}([-0.5,0.5])^{|\nodes|}$
        \STATE $f(\graph) = \{C_0, \ldots, C_k\}$, \text{with} $u \in C_i$
        \STATE Compute $\candlist{u}$ as defined in Eq.~\eqref{eq:candlist}
        \STATE $t \gets 1$, $C'_i \gets C_i$, $A'_u \gets A_u$ , $\graph' \gets \graph$, $\bm{b}_u^{(0)} \gets \bm{0}$
        \WHILE {$sim(C_i \setminus \{u\}, C'_i \setminus \{u\}) > \tau$ \textbf{and} $t \leq T$}
            \STATE $\hat{p} \gets \tanh(\hat{p} - \eta \nabla_{\hat{p}} \, \mathcal{L}(\hat{p}; A_u, \candlist{u}, q))$
            \STATE $p \gets \operatorname{threshold}(\hat{p})$
            \STATE $A'_u \gets \operatorname{clamp}(A_u + p)$
            \STATE $\bm{b}_u^{(t)} \gets$ changes in $A'_u$ up to step $t$
            \IF{$\|\bm{b}_u^{(t)} - \bm{b}_u^{(t-1)}\|_0 > 0$}
                \STATE Update $\graph'$ based on $\bm{b}_u^{(t)}$
                \STATE $f(\graph') \gets \{C'_0, \ldots, C'_r\}$, \text{with} $u \in C'_i$
            \ENDIF
            \IF{$\|\bm{b}_u^{(t)}\|_0 > \beta$}
                \STATE $\hat{p} \sim \mathcal{U}([-0.5,0.5])^{|\nodes|}$
            \ENDIF
            \STATE $t \gets t+1$
        \ENDWHILE
        \STATE \textbf{return} $\graph'$
    \end{algorithmic}
\end{algorithm}

\smallskip
\noindent \textbf{\textit{{Convergence Guarantee}.}}
The convergence of \method\ is rooted in the principles of gradient-based optimisation. The objective function, as defined in Eq. \eqref{eq:p_objective}, comprises two components, $ \ell_{\textit{hide}} $ and $ \ell_{\textit{dist}} $, 
both involving the $L^q$-norm, with $q \geq 1$ (see Eqs. \eqref{eq:l-hide} and \eqref{eq:l-dist}).
Therefore, both terms are convex with respect to $\hat{p}$, and their combination ensures that the first part of the objective is convex. However, the constraint on the number of modified edges for node $ u $, expressed as  $ ||\bm{b}_u||_0 \leq \beta $, introduces non-convexity due to the $L^0$-norm. This makes the overall optimisation problem NP-hard~\cite{nemirovskii1983problem}, necessitating a numerical approximation via stochastic gradient-based methods.
\\
From a theoretical standpoint, our method leverages the guarantees of gradient-based optimisation in non-convex settings. By using a sufficiently small learning rate ($\eta$), we can ensure convergence to a stationary point, which may correspond to either a local minimum or a saddle point. However, achieving global optimality is not guaranteed in non-convex problems. To mitigate this limitation, we employ a strategy of running the method multiple times with different random initialisations, a widely adopted approach.


\smallskip
\noindent \textbf{\textit{{Computational Complexity}.}}
We examine the computational complexity of our approach  to determine its feasibility for deployment in large-scale production environments.
In this analysis, we consider a graph with $|\nodes| = n$ nodes and $|\edges| = m$ edges. The computational cost of our method primarily hinges on two key operations outside the optimisation process, as other operations reduce to simple $O(n)$ vector computations. Let $F(n,m)$ represent the cost of applying the detection algorithm $\f$, and $\tilde{F}(n,m)$ denote the cost of constructing the vector $\candlist{u}$. 
With a maximum of $T$ iterations for the optimisation process, the total computational complexity is:
\begin{equation}
    O \bigg[ n + \tilde{F}(n,m) + F(n,m) + T ( n + F(n,m)) \bigg].
\end{equation}
In our implementation, the construction of $\candlist{u}$ is dominated by betweenness centrality calculations, leading to a complexity of $\tilde{F}(n,m) = O(mn)$. For the detection algorithm $\f$, the complexity depends on the specific community detection method employed. 


\subsection{\method{}-Projected}
The optimisation loop in Algorithm \ref{alg:pseudocode} does not guarantee that the counterfactual graph fully exhausts the available modification budget. In contrast, state-of-the-art methods typically enforce $\|\bm{b}_u\|_0 = \beta$ by applying changes until the budget is exactly met. To ensure that \method{} is not overly penalised in such comparisons, we introduce a \emph{projection step} performed after the optimisation loop, which adjusts the final perturbation to exactly match the prescribed budget.
Let $T^*$ be the iteration at which \method{} terminates. We define the used budget as $\beta_{\textit{used}} = \|\bm{b}^{(T^*)}_u\|_0$, and the remaining budget as $\beta_{\textit{rem}} = \beta - \beta_{\textit{used}}$. 
Starting from the final perturbation parameters $\hat{p}$ and the latest discrete adjacency vector $A_u'$, we derive a momentum-smoothed descent direction by computing an exponentially weighted average of the stored gradients $\{g^{(t)}\}_{t=1}^{T^*}$:
\begin{equation}
\bar{g} = (1-\gamma)\sum_{t=1}^{T^*} \gamma^{T^*-t} g^{(t)},
\end{equation}
where $\gamma$ is the decay factor controlling the weighting of past gradients, set to $0.9$ in accordance with the literature \cite{kingma2015iclr}. As long as $\beta_{\textit{rem}} > 0$, we update the parameters as $\hat{p} = \hat{p} - \eta\,\bar{g}$, then discretise and clamp the result to obtain a new binary adjacency vector $A_u'$.

Let $\delta$ denote the number of new modifications introduced in one projection step. When $0 < \delta \le \beta_{rem}$, we apply all proposed changes to the counterfactual graph $G'$, and update the remaining budget as $\beta_{\textit{{rem}}} = \beta_{\textit{rem}} - \delta$. Otherwise, if $\delta > \beta_{\textit{rem}}$, we score each candidate modification involving node $v$ (the node to be attached to or detached from 
$u$) by
$s_v = |\hat{p}_v| \cdot |\bar{g}_v|$, favouring changes with strong gradient support. We then sort candidates in descending order of $s_v$ and apply the top $\beta_{\textit{rem}}$ changes. This procedure guarantees the entire budget is exhausted ($\| \bm{b}_u\|_0 = \beta$), while keeping the final perturbation aligned with the optimiser's search direction. We refer to this variant of our method as \method{}-P.

\section{Experiments}
\label{sec:experiments}


\subsection{Experimental Setup}
\label{sec:exp-setup}

\textbf{\textit{Datasets.}} We evaluate our method on a diverse collection of real-world undirected graphs, encompassing a range of domains and data types. These include social and human interaction networks (\texttt{kar},\footnote{\label{fn:konect}\href{http://konect.cc/}{http://konect.cc/}}  Wikipedia's \texttt{vote},\footnote{\label{fn:networkrepo}\href{https://networkrepository.com}{https://networkrepository.com}} and Facebook \texttt{fb-75}\footref{fn:networkrepo}), information networks (\texttt{words}\footref{fn:konect} and \texttt{arxiv}\footnote{\label{fn:arxiv}\href{https://snap.stanford.edu/data/}{https://snap.stanford.edu/data/}} Condensed Matter), and infrastructure networks (US Power Grid \texttt{pow}\footref{fn:konect}).

\smallskip
\noindent \textbf{\textit{Community Detection Algorithms.}} 
We consider four community detection algorithms: two modularity-based approaches -- \textit{greedy}~\cite{greedy_detection_alg} and \textit{leiden}~\cite{leiden}; the \textit{walktrap} algorithm~\cite{walktrap_detection_alg}, which relies on random walks; and \textit{dgcluster}~\cite{deepl_dgcluster}, a deep learning-based method that performs clustering using node attributes -- in our case, embeddings are generated with \textit{node2vec}~\cite{node2vec}. 
\\
Table \ref{tab:datasets_and_communities} summarises the datasets used in our evaluation, including their size and the number of communities per algorithm.

\begin{table}[htpb]
\centering
\caption{Properties of the graph datasets considered in this work, including the number of communities identified by \textit{greedy}, \textit{leiden}, \textit{walktrap}, and \textit{dgcluster}.}
\label{tab:datasets_and_communities}
\scalebox{0.95}{
    \begin{tabular}{l c c c c c c }
        \toprule
        \multirow{2}{*}{Dataset} &  \multirow{2}{*} {$|\nodes|$} &  \multirow{2}{*} {$|\edges|$} & \multicolumn{4}{c}{Number of Communities} \\
        \cmidrule(lr){4-7}
        & & & \textit{greedy} & \textit{leiden} & \textit{walktrap} & \textit{dgcluster} \\
        \midrule
        \rowcolor[gray]{0.95} \texttt{kar} & 34 & 78 & 3 & 4 & 5 & 2\\
        \texttt{words} & 112 & 425 & 7 & 7 & 25 & 4 \\
        \rowcolor[gray]{0.95} \texttt{vote} & 889 & 2,900 & 12 & 8 & 42 & 5\\
        \texttt{pow} & 4,941 & 6,594 & 41 & 38 & 364 & 664 \\
        \rowcolor[gray]{0.95}\texttt{fb-75} & 6,386 & 217,662 & 16 & 13 & 349 & 9 \\
        \texttt{arxiv} & 23,133 & 93,497 & 270 & 56 & 2306 & 1322\\
        \bottomrule
    \end{tabular}
    }
\end{table}

\smallskip
\noindent \textbf{\textit{Similarity Metric.}} 
To determine the success of obscuring community memberships, we use Sørensen-Dice coefficient~\cite{metrics:Dice} as the similarity function $sim(\cdot,\cdot)$ in Algorithm~\ref{alg:pseudocode}. This metric 
measures similarity between two sets, ranging from $0$ (no similarity) to $1$ (high similarity). The objective is achieved if $sim(C_i \setminus \{u\}, C'_i \setminus \{u\}) \leq \tau$.


\begin{figure*}[htbp]
    \centering
    \subfloat[$\beta = \mu/2$.]{%
       \includegraphics[width=0.32\linewidth]{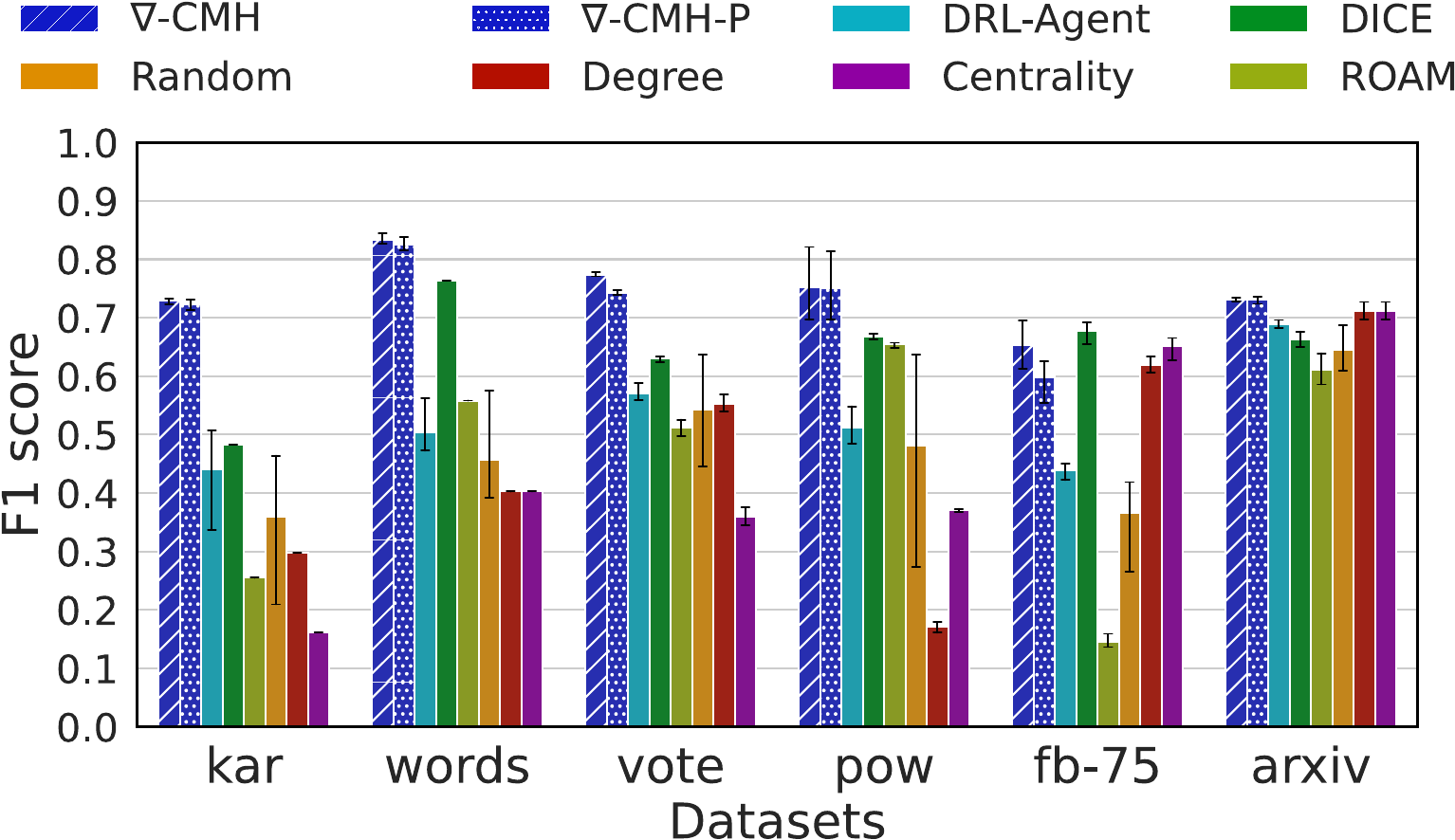}}
    \hspace{2mm}
    \subfloat[$\beta = \mu$.]{%
        \includegraphics[width=0.32\linewidth]{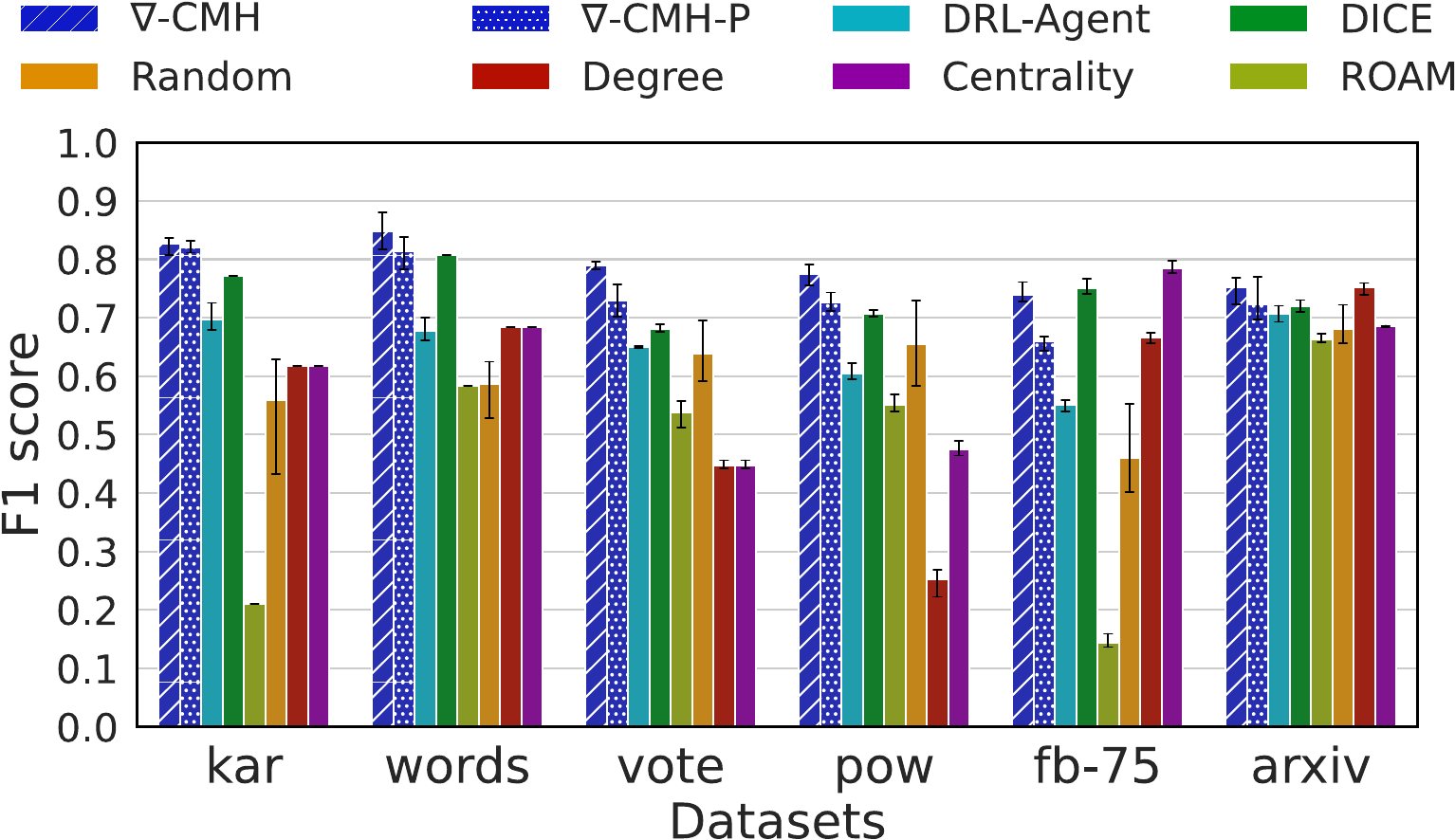}}
    \hspace{2mm}
    \subfloat[$\beta = 2\mu$.]{%
        \includegraphics[width=0.32\linewidth]{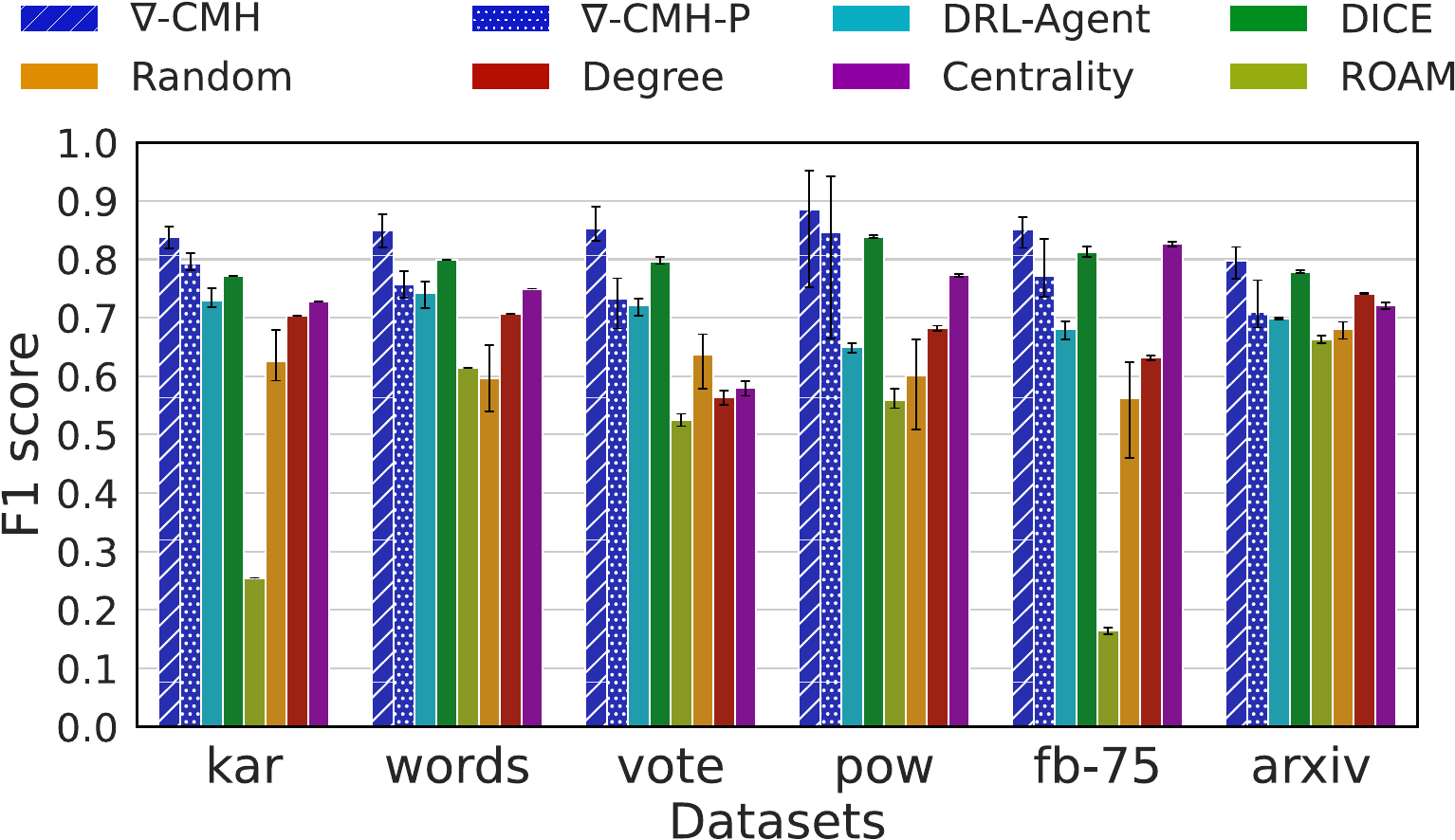}}
    \caption{F1 score between SR and NMI in the symmetric setting (\,$f(\cdot)$: \textit{greedy}) when $\tau=0.5$ across different budgets $\beta$.}
    \label{fig:f1-symmetric-greedy}
\end{figure*}

\smallskip
\noindent \textbf{\textit{Baselines.}} We compare the hiding assessment of our method, namely \method{}, against six baseline approaches:
\begin{enumerate}[itemsep=1pt, topsep=1pt]
    \item \textit{DRL-Agent}. A deep reinforcement learning method~\cite{bernini2024kdd}.
    \item \textit{DICE}. This heuristic, originally proposed for hiding communities, is based on the principle of \emph{Disconnect Internally, Connect Externally}~\cite{deception_modularity_2}. We adapt it to our setting by removing one edge of $u$ within its current community, specifically targeting the highest-degree node, and then creating $\beta-1$ external connection from $u$ with the same logic.
    \item \textit{ROAM}. This approach builds on the Roam heuristic~\cite{deception_modularity_2}, which is originally developed to  reduce a node’s centrality in a graph. It works by removing the connection to the highest-degree neighbour $v_0$, and then adding up to $\beta-1$ new edges from $v_0$ to neighbours of 
    $u$ that are not already connected to $v_0$, selecting them based on high-degree.
    \item \textit{Random-based}. This method picks a random node from $V$, and either adds or remove the connection to it. 
    \item \textit{Degree-based}. It modifies an edge that involves the highest-degree node, either by removing or adding the link.
    \item \textit{Centrality-based}. This approach alters the edge connected to the node having the highest betweenness centrality~\cite{freeman1977betweenness}. 
\end{enumerate}

\smallskip
\noindent \textbf{\textit{Evaluation Metrics.}}
We evaluate the effectiveness of each method in achieving the node hiding goal using the following metrics.
\begin{enumerate}[itemsep=1pt, topsep=1pt]
    \item \textit{Success Rate} (SR). It measures the percentage of cases in which the target node $u$ is successfully hidden from its original community, i.e., when $sim(C_i \setminus \{u\}, C'_i \setminus \{u\}) \leq \tau$. Higher values indicate better performance.
    \item  \textit{Normalised Mutual Information} (NMI). To assess the impact of the counterfactual graph $\graph'$ on the resulting community structure $f(\graph')$, we compute the NMI score~\cite{nmi_1,nmi_2} between $f(\graph')$ and the original structure $ f(\graph)$.
    Higher values indicate greater similarity, and thus a lower cost.
\end{enumerate}
In general, SR and NMI are inherently contrasting metrics -- higher SR often corresponds to lower NMI, and vice versa. To evaluate the trade-off between them, we compute their harmonic mean (i.e., F1 score) using the following formula: $ \frac{2 \times \text{SR} \times \text{NMI}}{\text{SR} + \text{NMI}}$.

\smallskip
\noindent \textbf{\textit{$\nabla$-\,CMH/$\nabla$-\,CMH-P.}} We use Adam optimiser to solve the objective of Eq. \eqref{eq:p_objective}, setting $q=2$ and leaving the exploration of alternative norms to future work. We perform a Bayesian hyperparameter search for $\lambda, T, \eta$, and $\{a_i\}_{i=1}^K$ to maximise the F1 score between SR and NMI. Further details are reported in Section \ref{subsec:param-analysis}.

\smallskip
\noindent \textbf{\textit{Evaluation Protocol.}}
We analyse \method{}/\method{}-P under various parameter configurations. Specifically, we vary the similarity constraint $\tau$ with values in $ \{0.3, 0.5, 0.8\}$, and test across different fixed budget values $\beta \in \{\mu/2, \mu, 2\mu\} $, where $ \mu = |\edges|/|\nodes|$ represents the average node degree.\footnote{For \texttt{kar} and \texttt{pow}, we specifically use $ \mu = \frac{|\edges|}{|\nodes|} + 1$.}
To ensure diversity in the evaluation, we select $3$ communities of varying sizes -- approximately $30\% $, $50\% $, and $80\% $ of the largest community. Within each, we randomly sample up to $100$ nodes as targets. This sampling strategy enables broad coverage across different community types. Experiments are conducted under two distinct setups: \emph{symmetric} and \emph{asymmetric}. In the symmetric setup, the same community detection algorithm $\f$ is used for both the optimisation and evaluation. In contrast, the asymmetric setup evaluates our method using a detection algorithm $g(\cdot)$, which differs from the optimisation algorithm $\f$. This design allows us to assess the \emph{transferability} of our method -- that is, its ability to generalise to detection algorithms it was not explicitly optimised against. Specifically, we use \textit{greedy} as $\f$ to ensure consistency with DRL-Agent's training setup. In the asymmetric setting, we use one of \textit{leiden}, \textit{walktrap}, or \textit{dgcluster} as $g(\cdot)$. To account for the inherent stochasticity in the detection algorithms, hiding methods, and sampling process, results are presented as the mean $\pm$ standard deviation computed over three independent runs.


\subsection{Results and Key Findings}
\label{subsec:results}

We evaluate \method{}/\method{}-P along three key axes. First, we assess its effectiveness in hiding target nodes from their communities, comparing against all baselines. Second, we evaluate its transferability in an asymmetric setting, where $\f$ differs from the actual community detection algorithm in use. Third, we analyse its computational efficiency relative to its main competitor, DRL-Agent.

\smallskip
\noindent \textbf{\textit{Hiding Assessment (Symmetric).}}
In Fig.~\ref{fig:f1-symmetric-greedy}, we compare the performance of \method{} with selected baselines in the symmetric setting. Specifically, we report the aggregated quality score -- computed as the F1 score between SR and NMI -- across varying budgets.
\\
Our method consistently outperforms all baseline approaches, except for a specific case on the \texttt{fb-75} dataset when $\beta = \mu/2$.
We attribute this to the structural characteristics of the graph, where heuristic methods such as Centrality or DICE tend to perform particularly well. Indeed, these approaches prioritise modifying edges connected to highly influential nodes, as discussed in Section~\ref{subsec:importance}.
\\
Moreover, it is worth noting that, unlike baseline methods that fully exhaust their allocated budgets, \method{} employs its resources more strategically and efficiently. As shown in Table~\ref{tab:used_budget_beta1}, which summarises budget usage across all datasets, our method achieves strong performance without utilising the entire budget. In fact, when baseline methods are constrained to the same budget used by \method{}, the performance gap becomes even more pronounced. This highlights the ability of our method to balance effectiveness with resource efficiency, achieving robust results at a lower cost. 
A similar trend holds across different values of $\tau$.

\begin{table}[htpb]
    \centering
    \caption{Average budget usage (absolute/\%) by \method{} across all datasets, fixed budget settings ($\beta \in \{\mu/2, \mu, 2\mu\}$), and $\tau=0.5$.}
    \label{tab:used_budget_beta1}
    \scalebox{0.88}{
    \begin{tabular}{ccccccc}
        \toprule
        \multirow{2}{*}{$\beta$} & \multicolumn{6}{c}{Dataset} \\
        \cmidrule(lr){2-7}
        & \texttt{kar} & \texttt{words} & \texttt{vote} & \texttt{pow} & \texttt{fb-75} & \texttt{arxiv}\\
        \midrule
        \rowcolor[gray]{0.95} $\mu/2$ & $1.0$\,/\,$100$ &
        $1.0$\,/\,$100$ & $1.0$\,/\,$100$ & $1.0$\,/\,$100$ &
        $11.3$\,/\,$66.3$ & $1.7$\,/\,$87.2$ \\
        $\mu$ & $2.3$\,/\,$76.7$ & $1.9$\,/\,$63.0$ & $2.2$\,/\,$71.7$
        & $1.8$\,/\,$90.8$ & $21.5$\,/\,$63.3$ & $3.1$\,/\,$78.2$ \\
        \rowcolor[gray]{0.95} $2\mu$ & $3.8$\,/\,$63.5$ &
        $3.4$\,/\,$56.2$ & $3.6$\,/\,$60.8$ & $2.9$\,/\,$71.5$ &
        $39.9$\,/\,$58.7$ & $5.2$\,/\,$65.4$ \\
        \bottomrule
    \end{tabular}
    }
\end{table}
Interestingly, the variant of our method that is forced to fully exhaust the available budget (\method{}-P) offers no performance gain. In fact, its additional modifications tend to degrade results, likely because these extra graph rewiring operations are redundant. Their associated cost outweighs any marginal benefit beyond what is already achieved by the more selective modifications of \method{}.



\begin{figure*}[htpb]
    \centering
    \subfloat[$g(\cdot)$: \textit{leiden}.]{%
       \includegraphics[width=0.32\linewidth]{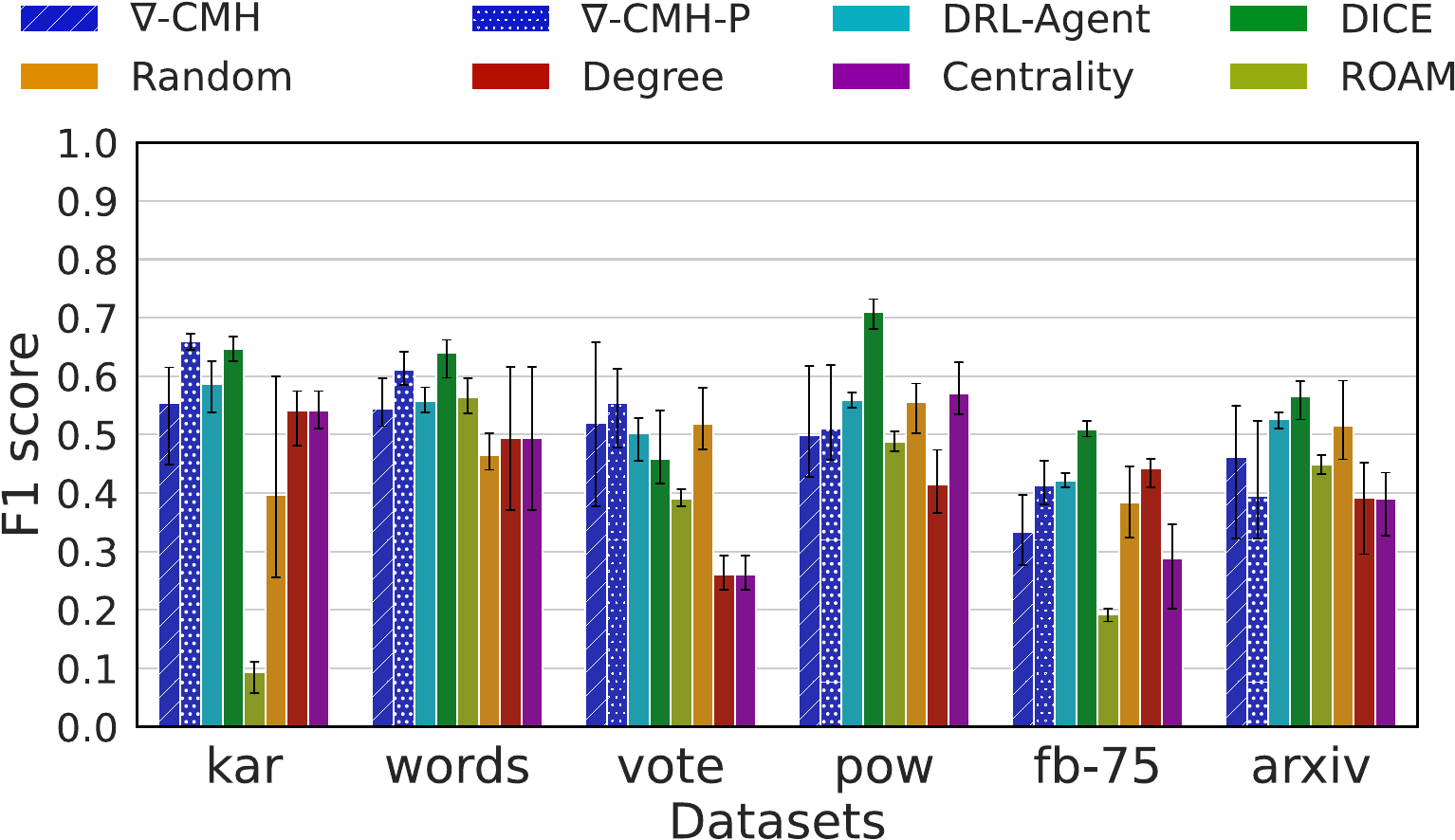}}
    \hspace{2mm}
    \subfloat[$g(\cdot)$: \textit{walktrap}.]{%
        \includegraphics[width=0.32\linewidth]{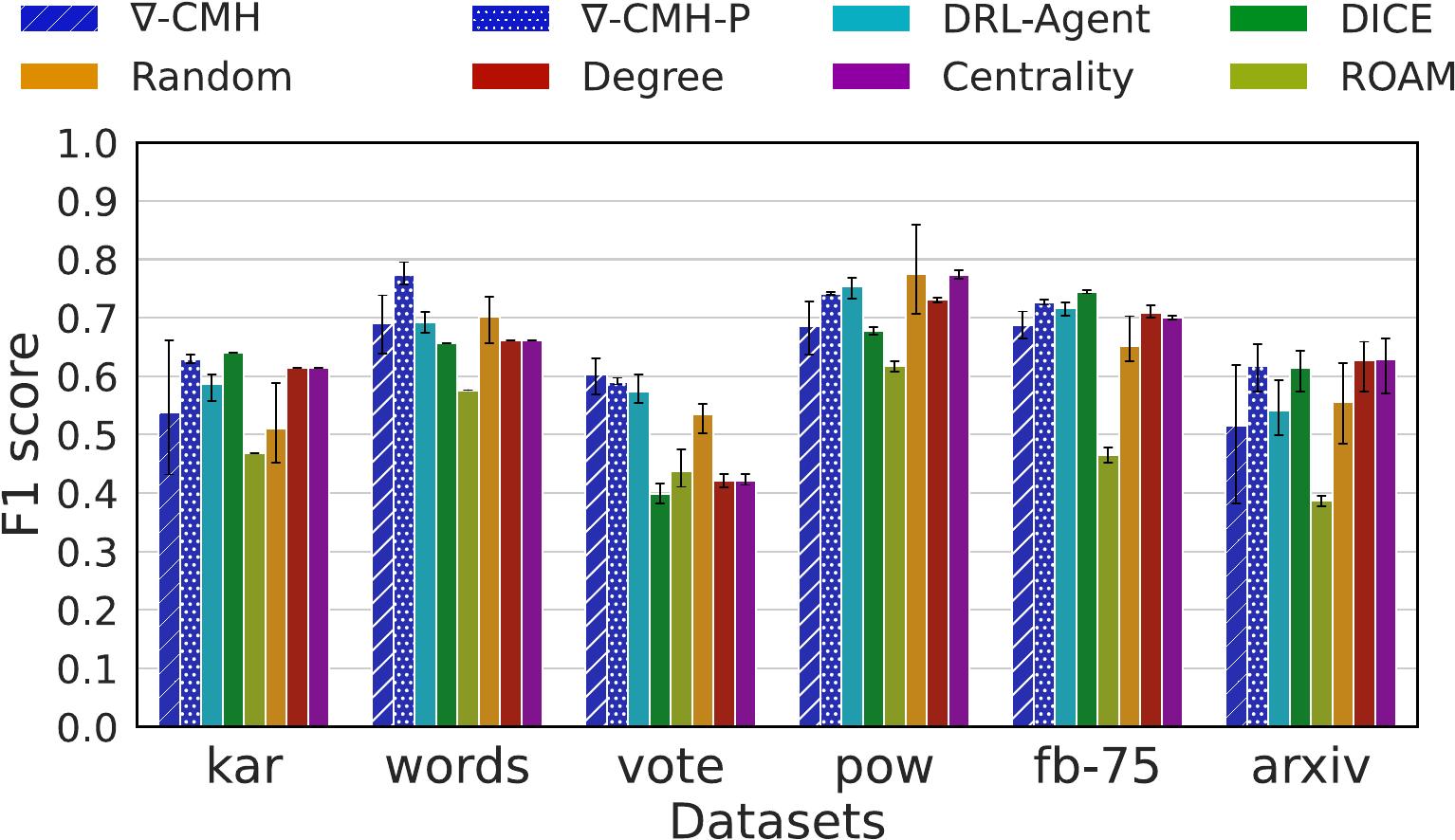}}
    \hspace{2mm}
    \subfloat[$g(\cdot)$: \textit{dgcluster}.]{%
        \includegraphics[width=0.32\linewidth]{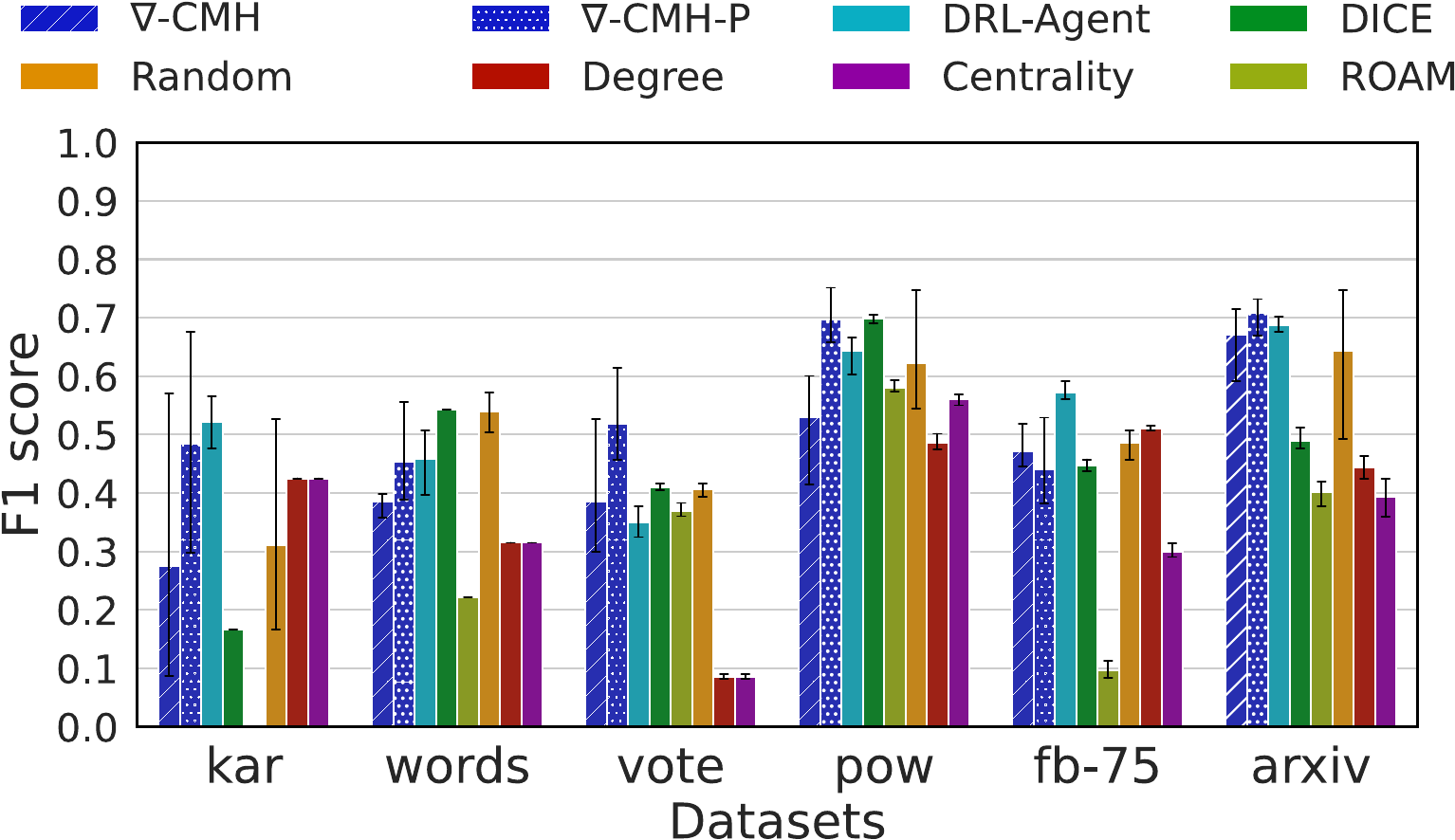}}
    \caption{F1 score between SR and NMI in the asymmetric settings ($f(\cdot)$: \textit{greedy}) when $\tau=0.5$ and $\beta=\mu$.}
    \label{fig:f1-asymmetric}
\end{figure*}

\smallskip
\noindent \textbf{\textit{Transferability (Asymmetric).}}
In Fig.~\ref{fig:f1-asymmetric}, we present the results for the asymmetric setting, which assesses each method's ability to generalise across different community detection algorithms. 
Our \method{} demonstrates transferability comparable to DRL-Agent, but generally lower than that of heuristic-based methods. This limitation is particularly evident on graphs such as \texttt{fb-75}, and remains consistent across different budget configurations. 
\\
We conjecture that this gap arises from a tendency of \method{} to ``overfit'' the specific community detection algorithm used during optimisation -- a limitation that heuristic-based methods inherently avoid due to their algorithm-agnostic nature. 
Notably, the variant \method{}-P, which is forced to exhaust the entire budget, suffers less from this overfitting tendency. It generally outperforms \method{} and DRL-Agent and, in some cases, even surpasses heuristic-based methods. We attribute this to its ability to explore a broader range of graph rewiring operations, which may help it uncover patterns that transfer more effectively to unseen algorithms, thereby improving generalisation.
Addressing this challenge opens a path for future improvements. In particular, reducing overfitting in asymmetric scenarios could involve optimising over multiple community detection algorithms rather than just a single one.

\smallskip
\noindent \textbf{\textit{Computational Efficiency.}}\footnote{All experiments were conducted on a GPU NVIDIA GeForce RTX 4090 and an AMD Ryzen 9 7900 12-Core CPU.}
Lastly, our method exhibits notable computational efficiency, operating faster than DRL-Agent. This speed advantage further enhances its practical applicability in real-world scenarios. Table \ref{tab:speedups} illustrates the average hiding time across datasets in the symmetric setting with $\tau=0.5$ and $\beta = \mu$, clearly showing that \method{} consistently outperforms the agent.

\begin{table}[H]
\centering
\caption{Average hiding time (secs.) of \method{} compared to DRL-Agent using $\f$: \textit{greedy}, with $\tau=0.5$ and $\beta=\mu$.}
\label{tab:speedups}
\scalebox{1}{
    \begin{tabular}{ l c c c c c }
        \toprule
        \multirow{2}{*} {Dataset} &  \multicolumn{2}{c}{Algorithm} & \multirow{2}{*} {Time Speed-up} \\
        \cmidrule(lr){2-3}
        & \method{} (ours) & DRL-Agent \\
        \midrule
        \rowcolor[gray]{0.95} \texttt{kar} & $0.013$ & $0.027$ & $\times 2.050 \ \blacktriangle$ \\
        \texttt{words} & $0.016$ & $0.047$ & $\times 2.932 \ \blacktriangle$ \\
        \rowcolor[gray]{0.95} \texttt{vote} & $0.063$ & $0.166$ & $\times 2.643 \ \blacktriangle$ \\
        \texttt{pow} & $0.115$ & $0.213$ & $\times 1.851 \ \blacktriangle$ \\
        \rowcolor[gray]{0.95}\texttt{fb-75} & $22.169$ & $110.463$ & $\times 4.983 \ \blacktriangle$ \\
        \texttt{arxiv} & $2.980$ & $5.227$ & $\times 1.754 \ \blacktriangle$ \\
        \bottomrule
    \end{tabular}
    }
\end{table}
To summarise, our experiments yield three key insights. First, in the symmetric setting, \method{} consistently outperforms all baselines. Fully exhausting the budget with \method{}-P offers no additional benefit, as \method{} already identifies the minimal set of graph rewiring operations required to achieve the hiding goal. Second, learning-based methods such as \method{} and DRL-Agent exhibit lower transferability compared to heuristic-based techniques, which are inherently agnostic to the specific community detection algorithm. Notably, \method{}-P alleviates this limitation by exploring a broader range of graph modifications. Third, \method{} is significantly more efficient, achieving much faster runtimes than the other learning-based competitor, DRL-Agent.


\subsection{Parameter and Hyperparameter Analysis}
\label{subsec:param-analysis}
Our method depends on two distinct sets of critical factors:
\begin{itemize}
    \item \textit{method parameters}, which define the operational constraints of the CMH problem: the similarity threshold $\tau$ and the modification budget $\beta$.
    \item \textit{hyperparameters}, which control the optimisation process of \method{}: the learning rate $\eta$, the regularisation strength $\lambda$, the number of maximum iterations $T$, and the coefficients associated with the $K=4$ promising actions, denoted as $\{a_i\}_{i=1}^K$. Specifically, $\omega_1$ (betweenness centrality) is associated with $a_1$, $\omega_2$ (degree) with $a_2$, $\omega_3$ (intra-community degree) with $a_3$, and $\omega_4$ (inter-community degree) with $a_4$.
\end{itemize}
Hyperparameters are tuned to maximise performance on a validation set, while method parameters are set according to the target use-case and later analysed to assess their effect on behaviour.
For clarity, we first report the optimisation of hyperparameters and then examine sensitivity to the method parameters.

\smallskip
\noindent{\textbf{\textit{Hyperparameter Optimisation.}}}
Table~\ref{tab:hyperparams} lists the hyperparameters selected via Bayesian optimisation for the setting $\beta=\mu$, with the objective of maximising the F1 score between SR and NMI across various datasets.
\begin{table}[htpb]
    \centering
    \caption{Hyperparameters of our method across all datasets for $\beta=\mu$ ($a_1$: betweenness centrality; $a_2$: degree; $a_3$: intra-community degree; $a_4$: inter-community degree).}
    \label{tab:hyperparams}
    \scalebox{1}{
    \begin{tabular}{ccccccccc}
        \toprule
        \multirow{2}{*}{Dataset} & \multicolumn{7}{c}{
        Hyperparameters} \\
        \cmidrule(lr){2-8}
        & $\eta$ & $\lambda$ & $T$ & $a_1$ & $a_2$ & $a_3$ & $a_4$ \\
        \midrule
        \rowcolor[gray]{0.95} \texttt{kar}
        & $0.079$ & $1.71$ & $120$ & $0.33$ & $0.20$ & $0.21$ & $0.24$\\
        \texttt{words}
        & $0.006$ & $0.04$ & $110$ & $0.16$ & $0.26$ & $0.34$ & $0.22$ \\
        \rowcolor[gray]{0.95} \texttt{vote}
        & $0.017$ & $0.37$ & $140$ & $0.48$ & $0.25$ & $0.01$ & $0.24$\\
        \texttt{pow}
        & $0.008$ & $18.1$ & $130$ & $0.05$ & $0.17$ & $0.41$ & $0.35$ \\
        \rowcolor[gray]{0.95} \texttt{fb-75}
        & $0.004$ & $0.15$ & $140$ & $0.29$ & $0.59$ & $0.09$ & $0.01$\\
        \texttt{arxiv}
        & $0.001$ & $17.2$ & $140$ & $0.40$ & $0.21$ & $0.05$ & $0.32$ \\
        \bottomrule
    \end{tabular}
    }
\end{table}



\smallskip
\noindent{\textbf{\textit{Sensitivity to Method Parameters.}}}
In Table \ref{tab:param_sens}, we illustrate how variations in $\tau$ and $\beta$ affect the F1 score of SR and NMI in the community membership hiding task, comparing our method against the three best baselines.
The reported results correspond to the symmetric setting on the \texttt{vote} dataset. As expected, increasing the threshold simplifies the achievement of the concealment goal, raising the F1 score for a fixed budget by imposing less strict requirements on membership hiding. In contrast, increasing the budget for a fixed threshold does not necessarily improve performance, as it permits more substantial modifications to the neighbourhood, which can further reduce the NMI score. 


\begin{table}[htpb]
    \centering
    \caption{Impact of $\tau$ and $\beta$ on the F1 score between SR and NMI in the symmetric setting (\,$\f$: \textit{greedy}) on the \texttt{vote} dataset for \method{}, DRL-Agent, DICE, and Random. Bold indicates the best (highest) value; underlined values are second best.}
    \label{tab:param_sens}
    \scalebox{0.9}{
    \begin{tabular}{ccccccccccc}
        \toprule
        \multirow{2}{*}{$\tau$} & \multirow{2}{*}{$\beta$} & \multicolumn{4}{c}{Algorithm} \\
        \cmidrule(lr){3-6}
        & & \method{}\,(ours) & DRL-Agent & DICE & Random  \\
        \midrule
        \multirow{3}{*}{0.3} 
        & $\mu/2$ & $\mathbf{0.72 \pm 0.01}$ & $0.45 \pm 0.01$ & $\underline{0.62 \pm 0.01}$ & $0.42 \pm 0.02$ \\
        & $\mu$ & $\mathbf{0.77 \pm 0.02}$ & $0.56 \pm 0.01$ & $\underline{0.65 \pm 0.01}$ & $0.57 \pm 0.05$ \\
         & $2\mu$ & $\mathbf{0.85 \pm 0.03}$ & $0.64 \pm 0.01$ & $\underline{0.71 \pm 0.01}$ & $0.53 \pm 0.01$ \\
        \midrule
        \multirow{3}{*}{0.5} 
        & $\mu/2$ & $\mathbf{0.77 \pm 0.00}$ & $0.57 \pm 0.01$ & $\underline{0.63 \pm 0.00}$ & $0.54 \pm 0.08$ \\
         & $\mu$ & $\mathbf{0.79 \pm 0.01}$ & $0.65 \pm 0.00$ & $\underline{0.68 \pm 0.01}$ & $0.64 \pm 0.04$ \\
        & $2\mu$ & $\mathbf{0.85 \pm 0.03}$ & $0.72 \pm 0.01$ & $\underline{0.80 \pm 0.01}$ & $0.64 \pm 0.04$ \\
        \midrule
         \multirow{3}{*}{0.8} 
        & $\mu/2$ & $\mathbf{0.83 \pm 0.01}$ & $\underline{0.75 \pm 0.02}$ & $\underline{0.75 \pm 0.01}$ & $0.71 \pm 0.04$ \\
        & $\mu$ & $\mathbf{0.86 \pm 0.01}$ & $0.80 \pm 0.01$ & $\underline{0.82 \pm 0.00}$ & $0.80 \pm 0.00$ \\
        & $2\mu$ & $\mathbf{0.90 \pm 0.04}$ & $0.78 \pm 0.01$ & $\underline{0.82 \pm 0.00}$ & $0.76 \pm 0.01$ \\
        \bottomrule
    \end{tabular}
    }
\end{table}

\vspace{-4mm}
\subsection{The Importance of Modified Edges} 
\label{subsec:importance}
So far, we have treated every edge modification uniformly, namely, adding or removing an edge has the same ``cost'' regardless of the other node involved in the modification. 
However, this assumption often does not hold in practice. For example, a user attempting to hide from a community may hesitate to remove or add a connection with a highly influential individual, while they may have no issue modifying connections with a less prominent one.

Heuristic-based methods like Centrality or DICE, though demonstrating strong transferability performance, achieve this by systematically prioritising modifications that affect edges connected to high-degree (i.e., important) nodes, by design.
This intrinsic bias can limit their practicality in real-world settings, where achieving the hiding goal should discourage from altering connections to key nodes.
In contrast, \method{} avoids this limitation by not restricting its modifications to edges involving high-importance nodes, making it more suitable for realistic scenarios.

To further validate this claim, we compute the average PageRank score~\cite{pagerank} of the nodes involved in the edge modifications performed by each method, across all datasets and target nodes considered.
In Table~\ref{tab:avg-pr}, we report the result of this experiment. We may observe that \method{} modifies connections with nodes having the smallest average PageRank, i.e., less important nodes. On the other hand, methods like DICE explicitly focus on influential nodes to achieve the hiding goal. 
This inherently biased behaviour highlights a fundamental limitation of heuristic-based techniques in practical applications, which \method{} overcomes.


\begin{table}[htpb]
\vspace{-2mm}
    \centering
    \caption{Average PageRank of the nodes involved in the perturbation, as computed by our method (\method{}) and the baselines (DICE and Centrality), for $\beta=\mu$. Bold indicates the best (lowest) value; underlined values are second best.
    Differences are all statistically significant ($\alpha=0.01$).
    }
    \label{tab:avg-pr}
    \scalebox{0.75}{
    \begin{tabular}{ccccccc}
        \toprule
        \multirow{2}{*}{Algorithm} & \multicolumn{6}{c}{Dataset} \\
        \cmidrule(lr){2-7}
        & \texttt{kar} & \texttt{words} & \texttt{vote} & \texttt{pow} & \texttt{fb-75} & \texttt{arxiv}\\
        \midrule
        \rowcolor[gray]{0.95} \method{} &
        $\mathbf{4.5 \cdot 10^{-2}}$ & $\mathbf{9.2 \cdot 10^{-3}}$ &
        $\mathbf{1.2 \cdot 10^{-3}}$ & $\mathbf{2.9 \cdot 10^{-4}}$ &
        $\mathbf{1.6 \cdot 10^{-4}}$ & $\mathbf{1.4 \cdot 10^{-4}}$ \\
        DICE & 
        $\underline{7.8 \cdot 10^{-2}}$ & $\underline{3.1 \cdot 10^{-2}}$ & $\underline{9.3 \cdot 10^{-3}}$ & $7.3 \cdot 10^{-4}$ & $7.8 \cdot 10^{-4}$ & $\underline{7.1 \cdot 10^{-4}}$\\
        \rowcolor[gray]{0.95} Centrality & 
        $8.9 \cdot 10^{-2}$ &
        $3.8 \cdot 10^{-2}$ & $1.2 \cdot 10^{-2}$ & $\underline{3.9 \cdot 10^{-4}}$ & $\underline{7.5 \cdot 10^{-4}}$ & $7.9 \cdot 10^{-4}$\\
        \bottomrule
    \end{tabular}
    }
\end{table}

\section{Current Limitations and Future Work}
\label{sec:limitations}
Like any approach tackling a new problem, \method{} has certain limitations that open avenues for future work. 

\smallskip
\noindent
\textbf{\textit{Scalability.}} 
While we have evaluated the scalability of \method{} on relatively large graphs such as \texttt{fb-75} and \texttt{arxiv}, and demonstrated its efficiency compared to its closest competitor (DRL-Agent), its applicability to truly large-scale graphs -- i.e., on the order of millions or billions of nodes -- though promising, needs further validation.

\smallskip
\noindent
\textbf{\textit{Hiding Goal.}} 
The notion of being ``masked'' from a community is operationalised in this work as a binary event, determined by whether the similarity between the original community and the one assigned after perturbation falls below a given threshold $\tau$. This abstraction, while practical, may not fully capture more nuanced or context-dependent definitions of community affiliation.

\smallskip
\noindent
\textbf{\textit{Side Effects.}}
Solving the community membership hiding task for a single node by modifying its local neighbourhood can unintentionally impact other nodes. 
For example, obscuring a node from its original community could reassign neighbouring nodes or alter the structure of nearby communities. 
In practice, we recommend post hoc validation to identify such side effects \textit{before} deployment. 
Nonetheless, we consider this assumption reasonable, since \method{} is primarily intended for use by entities controlling the graph network (e.g., online platforms), rather than individual users.

\smallskip
\noindent
\textbf{\textit{Multiple Nodes/Multiple Communities.}}
Our current formulation assumes a single target node requesting to be masked from a single community. In more realistic settings, multiple nodes may simultaneously seek to obscure their affiliations, possibly across several different communities. This scenario introduces new challenges, such as coordinating perturbations and mitigating compounded side effects, and remains an important direction for future work.

\vspace{-2mm}
\section{Conclusion}
\label{sec:conclusion}
We presented \method{}, a counterfactual graph generator designed to solve the community membership hiding task through gradient-based optimisation. This method employs a perturbation vector, added element-wise to the adjacency vector of the target node to mask. 
To ensure differentiability, we define an intermediate real-valued perturbation vector and a loss function that encourages minimal changes to the graph under consideration.
Graph modifications are further guided by a vector of \textit{promising actions}, namely the set of allowed operations to successfully escape the community. 

Experimental results demonstrate the superiority of \method{}, which consistently outperforms existing approaches in concealing target nodes. Moreover, our method efficiently identifies solutions that prioritise less disruptive graph rewiring operations, achieving strong performance without exhausting the allocated budget.
Finally, we introduced a variant, \method{}-P, which is forced to consume the entire budget. This variant improves transferability across different community detection algorithms, highlighting the potential for further enhancements in asymmetric scenarios.





\section*{Ethical Considerations}
\label{sec:ethical}
The \textit{community membership hiding} problem arises whenever a node in a generic graph must be concealed from a specific cluster of nodes. While broadly applicable in several domains, its most immediate and impactful use lies in safeguarding user privacy on online platforms, such as social networks.

In this section, we examine the potential ethical implications of community membership hiding techniques, using our method (\method{}) as a representative example. We assume that the platform owner (e.g., Meta, X, TikTok) offers \method{} as a service and executes it on behalf of users seeking to hide from a specific community (\textit{platform-mediated}). We do not address scenarios in which the end user performs this hiding independently (\textit{user-initiated}), as such cases involve a relaxed assumption about graph knowledge (i.e., shifting from global to local).

Although primarily designed to enhance individual privacy, \method{} possesses an inherent dual-use nature, making it essential to carefully weigh its potential benefits against its associated risks.

\smallskip
\noindent \textbf{\textit{Promoting Privacy and Autonomy.}}
At its core, community membership hiding aligns with fundamental privacy principles, including the ``right to be forgotten'' \cite{TRTBF2013,rosen2011right}, as recognised in regulations such as the European Union's GDPR \cite{regulation2016general} and AI Act \cite{act2024eu}. These frameworks emphasise individuals' control over their digital footprint and the prevention of unwanted inferences about sensitive personal information. 
\\
As community detection algorithms can inadvertently reveal political beliefs, health conditions, or affiliations with vulnerable groups, \method{} offers a crucial mechanism for digital autonomy. It provides a flexible and scalable alternative to the drastic measure of leaving online platforms entirely, allowing users to manage their visibility while retaining their online presence. For individuals in sensitive professions, such as journalists or human rights activists operating in authoritarian regimes, the ability to obscure their network affiliations could be vital for their safety and the integrity of their work. 
Similarly, it could combat online criminal activities by disrupting communication patterns among malicious users, though this application also carries its own set of ethical issues.

\smallskip
\noindent \textbf{\textit{Facing the Risks of Misuse.}}
While designed to enhance privacy, \method{} also has significant misuse potential. In the hands of malicious actors, this technique could obstruct legitimate graph analysis tools or security efforts by masking illicit or criminal activities within the network. Possible scenarios include:
\begin{itemize}
    \item \textit{Organised Crime}: Criminal groups obscuring their leadership structures and communication hubs to hinder law enforcement detection.
    \item \textit{Disinformation Campaigns}: State-sponsored or coordinated actors masking bot networks or inauthentic communities to evade platform moderation.
    \item \textit{Terrorist Recruitment}:  Extremist organisations hiding recruitment pathways and coordination channels from intelligence agencies.
\end{itemize}
This dual-use reality creates an inherent tension: while legitimate platforms might offer \method{} as a privacy-enhancing feature, malicious actors could exploit tools like that for illicit purposes. If widely available without safeguards, such tools could trigger an ``arms race'' in network obfuscation.

\smallskip
\noindent \textbf{\textit{Comprehensive Stakeholder Analysis.}}
A truly responsible ethical framework requires considering the diverse impacts on all affected stakeholders:
\begin{itemize}
    \item \textit{Target Users}: Gain privacy and control over their digital identity, safeguarding them from discrimination or harm.
    \item \textit{Non-Target Users/Communities}: May experience unintended ``side effects,'' where the modifications made to hide a target node inadvertently alter the community assignments or perceived affiliations of other, non-target individuals. This could lead to privacy breaches or mischaracterisations.
    \item \textit{Online Platforms}: Bear the responsibility of balancing user privacy with platform integrity and public safety. They must navigate complex legal and ethical landscapes, ensuring compliance with regulations while preventing misuse.
    \item \textit{Law Enforcement and Intelligence Agencies}: Face significant challenges if \method{} impedes their ability to investigate crimes, counter terrorism, or maintain public safety. This raises ethical dilemmas regarding the boundaries of individual privacy versus collective security.
    \item \textit{Policy Makers and Regulators}: Are tasked with developing legal and ethical frameworks that balance competing rights and interests in the digital space, potentially requiring new legislation to address the complexities of network privacy.
    \item \textit{Researchers and Developers}: Hold an ethical responsibility to consider the societal impact of their innovations. This includes responsible disclosure of potential vulnerabilities and contributing to the development of ethical guidelines for research in this sensitive area.
    \item \textit{Victims of Illicit Activities}: Could suffer direct and indirect harms if \method{} facilitates criminal or harmful online behavior, underscoring the societal cost of unchecked misuse.
\end{itemize}
\smallskip
\noindent \textbf{\textit{Mitigation Strategies and Safeguards.}}
Given these complex considerations, platforms offering community membership hiding capabilities must implement robust mitigation strategies and safeguards:
\begin{itemize}
    \item \textit{Access Controls and Permissions}: Hiding capabilities should not be granted indiscriminately. Platforms could implement tiered access based on factors such as user verification, content type, or compliance with strict terms of service. Each request to apply \method{} should be evaluated for its potential impact before being satisfied, possibly denying those associated with high-risk activities or likely misuse.
    \item \textit{Transparency Mechanisms}: While avoiding the revelation of exploitable vulnerabilities, platforms should be transparent about the general principles of \method{} implementation and usage. This could involve publishing aggregate statistics on hiding requests or clear, accessible user policies.
    \item \textit{Auditing and Monitoring}: Robust technical and organisational mechanisms are essential to detect and prevent the misuse of \method{}. This might include anomaly detection systems that flag unusual patterns of graph modifications or post-hoc analysis to identify potential illicit activities.
    \item \textit{Ethical Guidelines and Policies}: Platforms should develop internal ethical guidelines and comprehensive policies for the responsible deployment and governance of \method{} features, involving interdisciplinary teams (e.g., ethicists, legal experts, engineers).
    \item \textit{Collaboration with Law Enforcement}: Establishing clear, ethically sound protocols for collaboration with law enforcement agencies is crucial. This involves defining the conditions under which information might be shared, respecting legal due process, and protecting user rights.
\end{itemize}
\smallskip
\noindent \textbf{\textit{The ``Right to Hide'' vs. ``Right to Know'' Trade-off.}}
The core ethical tension inherent in \method{} lies in the conflict between an individual's ``right to hide'' their affiliations for privacy and society's ``right to know'' for public safety, law enforcement, and even legitimate academic research into social dynamics. This is not merely a problem of ``misuse'' but a fundamental clash of legitimate interests. When these two rights conflict, society must engage in a dialogue to establish clear boundaries, mechanisms for arbitration, and potentially new legal frameworks that address situations where individual privacy and collective security needs diverge.

\smallskip
\noindent \textbf{\textit{Alignment with Ethical AI Principles.}}
\method{}'s design also aligns with broader ethical AI principles:
\begin{itemize}
    \item \textit{Fairness}: Ensuring that the ability to hide community membership is equitably accessible and does not inadvertently create unfair advantages or disadvantages for certain users or groups.
    \item \textit{Accountability}: Establishing clear lines of responsibility for the design, deployment, and monitoring of \method{} systems, ensuring that mechanisms are in place for redress in cases of misuse or unintended harm.
    \item \textit{Transparency}: The gradient-based nature of \method{} offers a degree of interpretability, allowing for insights into why specific modifications are chosen. This intrinsic transparency is crucial for building user trust, enabling auditing, and distinguishing legitimate privacy-preserving actions from malicious obfuscation.
    \item \textit{Privacy by Design}: \method{} embodies the principle of ``privacy by design,'' integrating privacy considerations into the system's architecture from the outset, rather than as an afterthought.
\end{itemize}
In conclusion, while \method{} represents a significant technical advancement in social graph privacy, its deployment requires a profound and ongoing ethical commitment. Future research must not only advance the technical capabilities of such methods but also actively engage with the complex societal implications, fostering interdisciplinary dialogue to ensure responsible innovation that truly serves the public good.

\balance
\bibliographystyle{ACM-Reference-Format}
\bibliography{references}

\nocite{locale}

\end{document}